\documentclass[pra,aps,nobalancelastpage,twocolumn,superscriptaddress,longbibliography]{revtex4-1} 

\pdfoutput=1
\usepackage{showlabels}
\usepackage{color,amsthm,amsmath,amsxtra,amsfonts,dsfont,graphicx,bm}
\usepackage{booktabs,multirow}
\usepackage[section]{placeins}
\usepackage{float} 
\usepackage{etoolbox} 

\newcommand{\be}{\begin{equation}}
\newcommand{\ee}{\end{equation}}
\newcommand{\bea}{\begin{eqnarray}}
\newcommand{\eea}{\end{eqnarray}}

\newcommand\nn{\mathbf{n}}
\newcommand\ii{\mathbf{i}}
\newcommand\jj{\mathbf{j}}

\newcommand\mm{\mathbf{m}}
\newcommand\sss{\mathbf{s}}

\newcommand\kk{\mathbf{k}}
\newcommand\qq{\mathbf{q}}
\newcommand\pp{\mathbf{p}}

\newcommand\dd{\mathbf{d}}

\newcommand\rr{\mathbf{r}}
\newcommand\ttt{\mathbf{t}}

\newcommand\Ry {\text{Ry}}

\newcommand\hc{\text{H.c.}}

\newcommand{\ket}[1]{\left| #1\right\rangle}
\newcommand{\lla}[1]{\left\{ #1\right \} }

\newcommand{\pa}[1]{\left( #1\right)}
\newcommand{\ha}{H}

\newcommand{\abs}[1]{\left| #1\right|}

\newcommand{\fl}[1]{\left\lfloor #1\right\rfloor}

\newcommand{\st}[1]{_\text{#1}}

\usepackage{accents}

\usepackage{hyperref}

\definecolor{mygrey}{rgb}{0,0.45,0.45}
\definecolor{myblue}{rgb}{0.2,0.2,0.8}
\definecolor{myzard}{cmyk}{0,0,0.05,0}
\definecolor{mywhite}{rgb}{1,1,1}
\definecolor{mywhite}{rgb}{1,1,1}
\definecolor{myred}{rgb}{1,0.,0.3}

\newcommand{\ord}[1]{\mathcal{O}\pa{#1}}
\newcommand{\norm}[1]{| #1 |}
\newcommand{\co}[1]{\left[ #1\right]}
\newcommand{\MPQ}{Max-Planck-Institut f{\"u}r Quantenoptik, Hans-Kopfermann-Stra{\ss}e\ 1, D-85748 Garching, Germany}

\hypersetup{
  pdfsubject={},
  pdfcreator={pdflatex},
  colorlinks,
  linkcolor=blue,
  citecolor=blue,
  urlcolor=blue
}

\begin{document}

\title{Quantum Simulation of 2D Quantum Chemistry  in Optical Lattices
}

\date{\today}
\begin{abstract}
Benchmarking numerical methods in quantum chemistry is one of the key opportunities that quantum simulators can offer. Here,
we propose an analog simulator for discrete 2D quantum chemistry models based on cold atoms in optical lattices. We first analyze how to simulate simple models, like the discrete versions of H and H$_2^+$, using a single fermionic atom. We then show that a single bosonic atom can mediate an effective Coulomb repulsion between two fermions, leading to the analog of molecular Hydrogen in two dimensions. We extend this approach to larger systems by introducing as many mediating atoms as fermions, and derive the effective  repulsion law. In all cases, we analyze how the continuous limit is approached for increasing optical lattice sizes.
\end{abstract}

\author{Javier Arg\"uello-Luengo}
\email{javier.arguello@icfo.eu}
\affiliation{ICFO-Institut de Ci\`encies Fot\`oniques, The Barcelona Institute of Science and Technology,
08860 Castelldefels (Barcelona), Spain}

\author{Alejandro Gonz\'alez-Tudela}
\email{a.gonzalez.tudela@csic.es}
\affiliation{Instituto de F\'isica Fundamental IFF-CSIC, Calle Serrano 113b, Madrid 28006, Spain}

\author{Tao Shi}
\email{tshi@itp.ac.cn}
\affiliation{CAS Key Laboratory of Theoretical Physics, Institute of Theoretical Physics, Chinese Academy of Sciences, P.O. Box 2735, Beijing 100190, China}

\author{Peter Zoller}
\affiliation{Center for Quantum Physics, University of Innsbruck, A-6020 Innsbruck, Austria}
\affiliation{Institute for Quantum Optics and Quantum Information of the Austrian Academy of Sciences, Innsbruck, Austria.}

\author{J. Ignacio Cirac}
\email{ignacio.cirac@mpq.mpg.de}
\affiliation{\MPQ}
\affiliation{Munich Center for Quantum Science and Technology (MCQST), M\"unchen, Germany}
\maketitle

The field of theoretical quantum chemistry has experienced an
extraordinary progress due, in part, to many advances
in computational methods~\cite{szabo2012modern}. For instance,
Density Functional Theory~\cite{Hohenberg1964,Parr1989} has enabled
a better description and understanding of both static~\cite{Tsipis2014,headGordon,Alexandrova2006,Domingo2016}
and dynamic~\cite{Gross1990} properties of a large
variety of molecules. The capability of such computational
methods, whose main challenge is to address electronic correlations, are however sometimes hard to assess experimentally.
One approach is to use another (classical) computational technique that is exact in some restricted conditions, but can deal with large systems where exact calculations were not possible. The most prominent example is DMRG~\cite{White1992} which, despite the fact that it operates in 1D lattice systems, offers an ideal platform to benchmark DFT methods~\cite{Yang2019,Motta2019,Motta2017,Lubasch2016}.
 In more general scenarios, the field of quantum computing~\cite{Cao2019,Aspuru-2005,Lanyon2010,Kassal2011,Wecker2015,Higgott2019} can play a key role to overcome numerical limitations in the long-term, offering an
excellent setup to benchmark quantum chemistry computational methods.
Recently, we have proposed the alternative approach of analog quantum simulation~\cite{arguello2019analogue}, based on
the experimentally mature field of ultra-cold atoms~\cite{bloch08a,Esslinger2010,Gross2017}, where fermionic atoms play the role of the electrons. While quantum computers and analog simulators would certainly help to push quantum chemistry, the exploration of their full potentiality requires the development of techniques that go beyond the state of the art. 

In this Letter we propose and analyze a scheme for \emph{analog quantum chemistry simulation} that can be implemented with present technology. Our approach uses ultracold atoms to address lattice models in two spatial dimensions (2D), where the electron-electron interaction takes different forms. While not exactly reproducing all aspects of the real quantum chemistry scenario, this simulator still retains the most relevant ingredients, enabling the observation of the most representative phenomena in quantum chemistry. Furthermore, it offers a suitable platform to benchmark computational methods in that field. In particular, it allows us to extend the benchmarking offered by DMRG beyond 1D
~\footnote{We acknowledge that other analog simulators based on fermionic atoms trapped in optical lattices have been proposed to emulate the molecular potentials of benzene-like molecules~\cite{Luhmann2015} or simulate ultrafast dynamics in strong-fields~\cite{Sala2017,Senaratne2018}. 
In contrast to them, Ref.~\cite{arguello2019analogue} and the present proposal allow to go beyond the local interactions naturally found in cold atoms, simulating the non-local fermionic repulsion that appears in typical quantum chemistry problems.}.

For the sake of clarity, we will discuss
several scenarios, with increasing experimental difficulty,
for the simulation of quantum chemistry problems
in 2D discrete lattices that could later be compared to contemporary theoretical lattice methods, such as DFT or DMRG. We start
with simple one-electron systems, the analogous to the
Hydrogen atom, and the $H_2^+$ molecule. Then, we show
how to simulate two electron problems, here exemplified
by the $H_2$ molecule. Finally, we show how the system can
be scaled-up to more electrons, although with a different
dependence of the repulsion with the distance.

\emph{Model.}
In the following, we will consider a discrete version of quantum chemistry models in 2D. First, we start by considering a 2D square optical lattice of size $N\times N$. $N_f$ fermionic atoms, playing the role of electrons, can localize within the local minima of this optical lattice, and hop with nearest-neighbor tunneling rate $t_F$. The Hamiltonian describing their dynamics is then given by:
\begin{equation}
    \label{eq:hamKin}
    \ha_{K}=-t_F\,\sum_{\langle \ii, \jj \rangle}f^\dagger_\ii f_ \jj \,, 
\end{equation}
where $f_\ii^\dagger$ and $f_\ii$, are the creation and annihilation operators for a fermionic atom in the $\ii$-th lattice site~\footnote{Throughout the text, bold variables denote 2D vectors.
}, each of them separated by a lattice spacing $a$, and where the sum is taken over all nearest-neighbor pairs of lattice sites.
Fermionic atoms are subject to an external potential that induces the attraction to $N_{\text{nuc}}$ nuclei that we consider placed in fixed positions $\lla{\rr_n}_{n=1\ldots N_{\text{nuc}}}$~\footnote{In order to prevent the divergence in the origin, positions $r_n$ of the nuclei are shifted half a site from the lattice nodes in the $y$ direction.} (Born-Oppenheimer approximation~\footnote{Considering that the electronic dynamics is much faster than the nuclear one, their equations can be decoupled (Born-Oppenheimer approximation). The position $\lla{\tilde\rr_n}_{i=n\ldots N_n}$ of the $N_n$ nuclei is considered  fixed during the calculation of the electronic Hamiltonian $H_\text{cont}$, for the $N_f$ electrons in positions $\lla{\rr_i}_{i=1\ldots N_f}$.
\begin{equation*}
\begin{split}
       H_\text{cont}=&-\sum_{i=1}^{N_f}\frac{\hbar^2}{2m_e}\nabla^2_i -\sum_{i=1}^{N_f} \frac{1}{2}\sum_{n=1}^{N_n} Z_n V(\abs{\rr_i-\tilde\rr_n})\\
      & + \sum_{i\neq j=1}^{N_f}  V(\abs{\rr_i-\rr_j}) \,,
\end{split}
\end{equation*}
where $m_e$ is the mass of the electron and $Z_n$ is the atomic number of nucleus $n$. The first term then describes the kinetic energy of the electrons, the second its nuclear attraction following the potential $V(r)\,,$ and the third the electronic repulsion.
}),
\begin{equation}
\label{eq:hamNuc}
    H_{\text{n}}(\lla{\rr_n})=-\sum_{n=1}^{N_{\text{nuc}}}\sum_\jj Z_n V(\abs{\jj-\rr_n}) f^\dagger_\ii f_ \jj\,,
\end{equation}
where $Z_n$ is the atomic number of nucleus $n$, and $V(r)$ is the attractive nuclear potential~\footnote{This externally induced potential could eventually mimic the effect of inner-shell electrons as well.}.
In 2D lattices, this potential can be obtained by combining the light shift induced by an external laser orthogonal to the lattice and a fully programmable intensity mask using, for example, a digital mirror device~\cite{choi16a}.
Depending on the model to be simulated, we will also consider the Hamiltonian $H\st{med}$ describing a set of bosonic atoms that mediates fermion-fermion interactions according to some effective potential, $V\st{eff}$.

\begin{figure}[tbp]
\centering
 \includegraphics[width=1\linewidth]{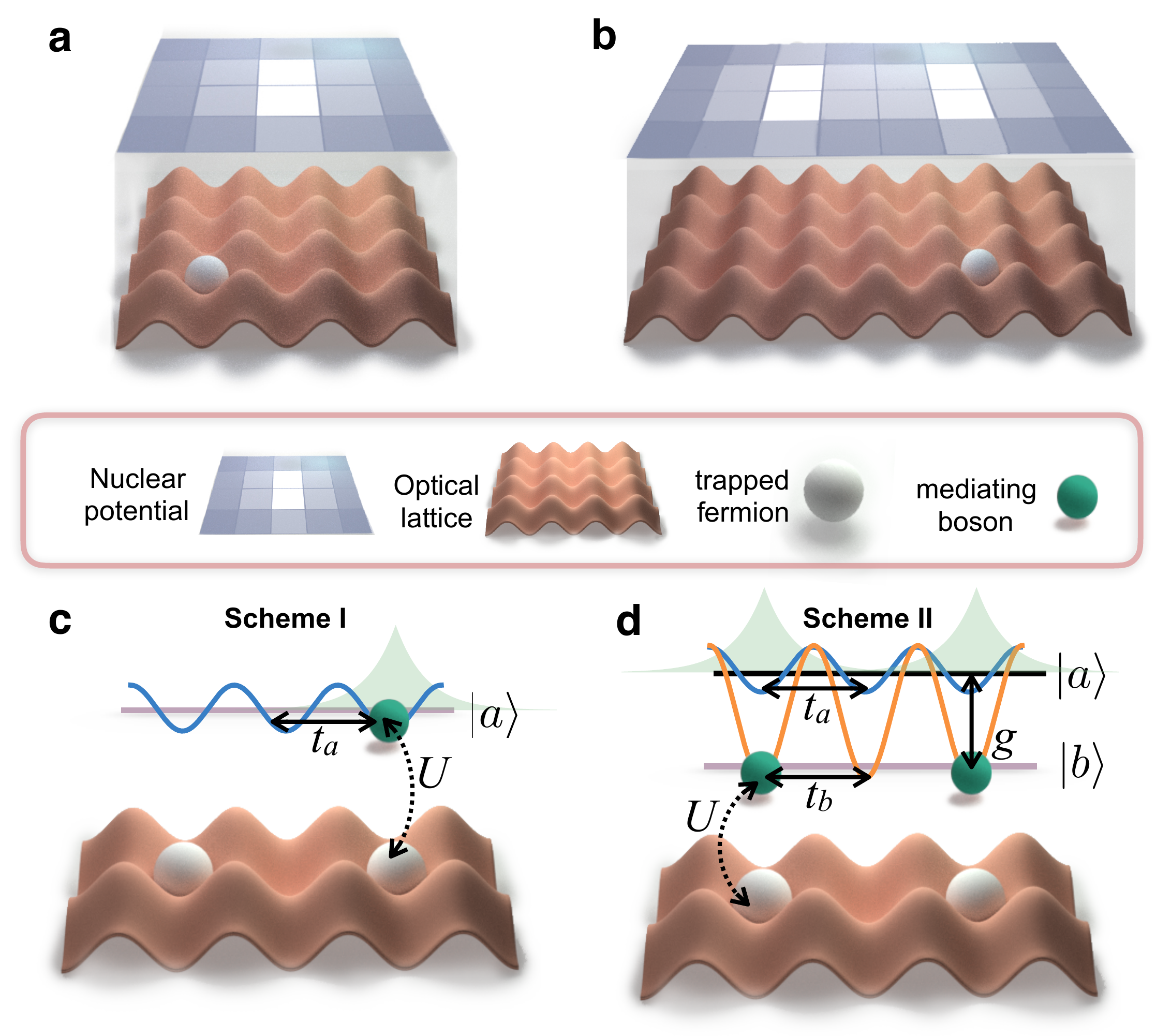}
\caption{Fermionic atoms (white) play the role of the molecular electrons. They hop in a 2D lattice (red), where the nuclear potential is imprinted (blue). For a single simulated electron, this pattern can lead to, e.g.,  atomic Hydrogen ((a), one nucleus) or $H_2^+$ ((b), two nuclei). For more than one fermionic atom, two different schemes are proposed to mediate an effective repulsion between them.
 (c) A single atom (green) is used. It tunnels with constant $t_a$ through a lattice with the same spacing as the fermionic one. There is an on-site repulsion with strength $U$ when the mediating atom occupies the same site as the fermion.  (d) We use as many mediating atoms as electrons need to be simulated (2 in the case of the figure). The on-site repulsion with the fermions now appears in a different internal level $b$, whose tunneling is slower as compared to level $a$, using a state-dependent lattice. Both levels are coherently coupled with coupling constant $g$.}
\label{fig:scheme}
\end{figure}

We consider now the simplest situation of simulating atomic Hydrogen. By choosing a potential with a unique nucleus $Z_1=1$ centered in the lattice site $\rr_1=(\lfloor N/2 \rfloor, \lfloor N/2 \rfloor+1/2)$, the total Hamiltonian reads as,
\begin{equation}
\label{eq:Hsimul}
    H_1=H_K+H_{n}(\rr_1)\,.
\end{equation} 
To begin with, we consider the attractive Coulomb potential on its standard form, $V(r)=V_0/r$, for moderate finite lattice sizes, e.g. $N=40$. In order to gain intuition, one can compare this discretized Hamiltonian to the continuum limit, where an analytical solution is also known in 2D~\cite{Zaslow1967}. As a consequence of the reduced dimensionality, electrons get closer to the nuclei than in the 3D case~\cite{Zhu1990}. Each energy level corresponds to $E_n^*=\frac{-\Ry}{(n-1/2)^2}$, for $n=1,2,\ldots$ In that limit, one can also identify,
\begin{equation}
a_0/a=t_F/V_0 \quad \text{ and } \quad \Ry =V_0^2/t_F\,,
\end{equation}
that are the equivalent Bohr radius $(a_0)$, and Rydberg energy $(\Ry)$, for the 2D discrete model~\footnote{As compared to the three-dimensional case, $\text{Ry\,(2D)}=\text{4Ry\,(3D)}$, and $2a_0\,\text{(2D)}=a_0\, \text{(3D)}$. Throughout the text, we will omit the (2D) labelling.}. The first ultimately determines the size of the orbitals and thus how the continuum limit is recovered. In particular, it is needed that the orbitals fit in the lattice (to avoid finite size effects), and that this Bohr radius occupies several lattice sites (to avoid discretization errors), leading to the inequalities,
\begin{equation}
	N \gg t_F/V_0 \gg 1\,.
\end{equation}

\begin{figure}[tbp]
\centering
 \includegraphics[width=1\linewidth]{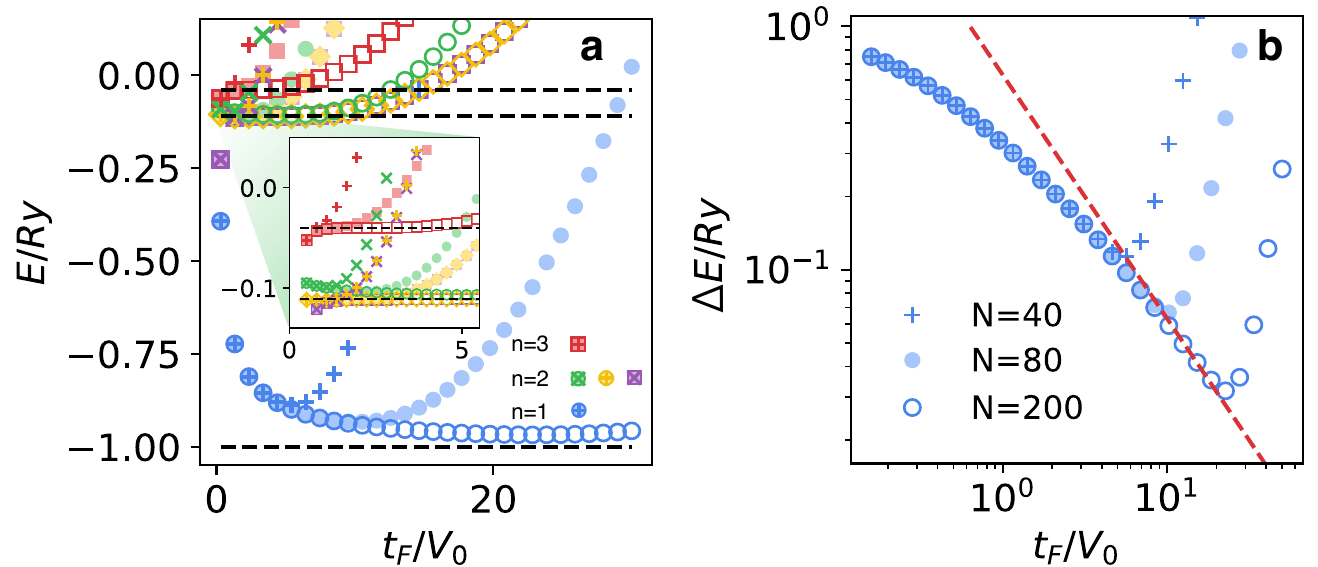}
\caption{ (a) Lower part of the spectrum for the discretized 2D atomic Hydrogen Hamiltonian in Eq.~\eqref{eq:Hsimul} for different values of the effective Bohr radius $t_F/V_0$. As more lattice sites are involved in the simulation ($t_F/V_0$ increases), the spectrum approaches the value in the continuum (horizontal lines for $n=1,2,3$). This is valid up to a critical Bohr-radius in which finite-size effects become relevant and the solution deviates from this behaviour. This critical value appears earlier for smaller sizes ($N=40$ for crossed markers) than for bigger systems ($N=80$, coloured marker, and $N=200$, edged marker). (b) The energy difference $\Delta E$ between the ground-state of the discretized Hamiltonian in Eq.~\eqref{eq:Hsimul}, and the one in the continuum decreases polynomially before finite-size effects become relevant~\cite{Note7}. Larger system sizes can follow this scaling up to more precise solutions. Dashed line follows the scaling $(t_F/V_0)^{-1}$. }
\label{fig:4Hydrogen2D_ED_atom1}
\end{figure}

In Fig \ref{fig:4Hydrogen2D_ED_atom1}(a) we show the lower part of the spectrum of the discretized Hamiltonian \eqref{eq:Hsimul} for different values of $t_F/V_0$ and $N$. First, we observe that we have quantized levels, and thus the discrete model qualitatively reproduces the continuous one. In fact, this can be observed with small lattices ($N=40$). Quantitatively, we see that by increasing the ratio $t_F/V_0$  and making the lattice larger, one approaches the continuum limit, as intuitively expected. The error for this approximation as a function of $t_F/V_0$ is shown in Fig. \ref{fig:4Hydrogen2D_ED_atom1}(b), where it is observed to scale approximately as $\pa{t_F/V_0}^{-1}$~\footnote{See the Supplementary material accompanying this Letter. Section A discusses the scaling of the spectrum of the discretized 2D Hamiltonian as the lattice size increases. Section B derives the effective interaction mediated by a single boson with one long-lived state. Section C focuses on the effective interaction mediated by several mediating atoms with two long-lived internal states. Section D includes further details about the numerical calculations shown in Fig. 2-4.}.

\begin{figure}[tbp]
\centering
\includegraphics[width=0.95\linewidth]{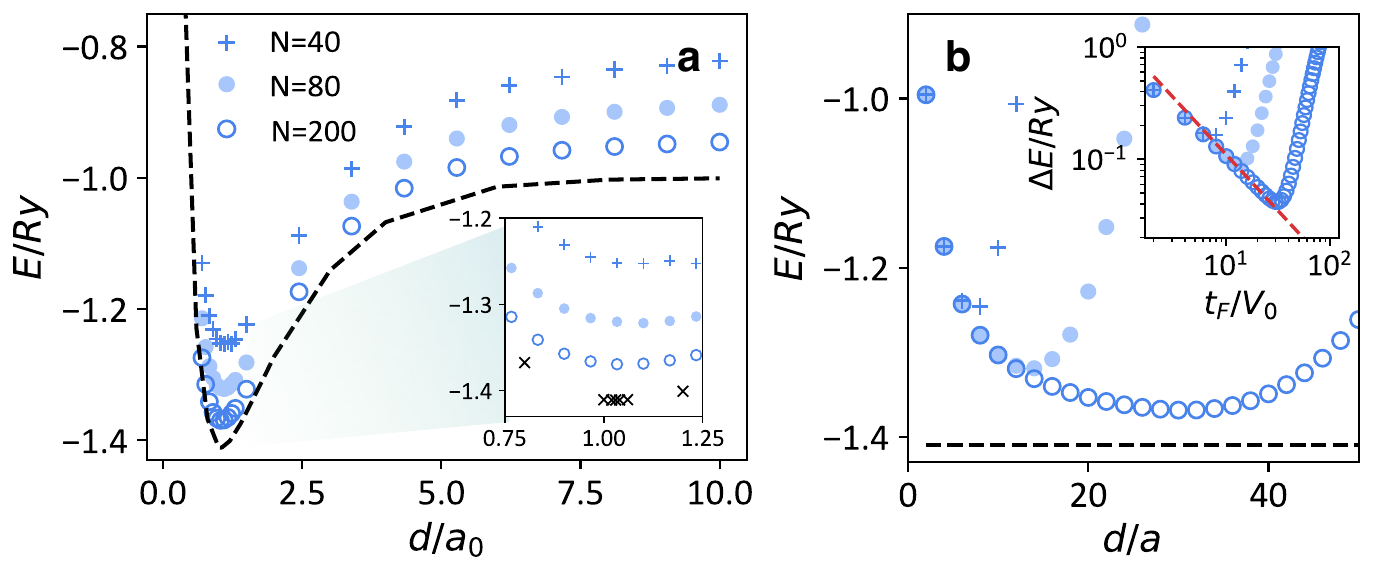}
\caption{ (a) Ground-state energy of the 2D hydrogen cation ($H_2^+$) for different lattice sizes $N$ and internuclear distance $d/a_0$ (see Text for the optimal choice of the lattice separation). The inset zooms into separation close to equilibrium. Dashed line (black crosses in the inset) follows an accurate solution for this 2D cation \cite{Patil2003}. (b) Ground-state energy of $H_2^+$ calculated for fixed $d/a_0=1$ and increasing effective Bohr radius $t_F/V_0$. The solution decreases up to a critical size at which finite-size effects appear. This critical size is larger for bigger lattice sizes. In the inset, the difference in energies to the tabulated value $-1.41\,\Ry$ (black dashed line) reveals the scaling $(t_F/V_0)^{-1}$ (red dashed line). Markers represent the same sizes as in (a).}
\label{fig:Hmas}
\end{figure}

Let us now explore a system with a single fermion and two equal nuclei, $Z_{1,2}=1$, separated by $d/a$ lattice sites, $\rr_{1,2}=(\lfloor N/2\pm d/(2a) \rfloor, \lfloor N/2 \rfloor+1/2)$, i.e. the analog of $H_2^+$. This internuclear separation measured in number of lattice sites can be directly expressed in terms of the Bohr radius as  $d/a_0=(d/a)\cdot (V_0/t_F)$, and therefore compared to tabulated values~\cite{Patil2003}. 
In Fig.~\ref{fig:Hmas}(a) we plot the energy of the ground state as a function of the distance. We obtain a molecular potential, as it is expected for $H_2^+$, already for the moderate size $N=40$. 
Increasing $t_F/V_0$ favors accuracy, up to the point where finite-size effects appear. At this point the difference in energies to the continuum (dashed line) deviates from the universal scaling $\Delta E \propto \pa{t_f/V_0}^{-1}$, which identifies the optimal configuration for our finite system and a given choice of $d/a_0$. In Fig.~\ref{fig:Hmas}(b) we illustrate this effect by showing that a given internuclear separation $d/a_0$, can be calculated with different values of integer lattice-site separations $d/a$ by tuning the effective Bohr radius $a_0/a$ accordingly (see \cite{Note7}). 

\begin{figure}[tbp]
\centering
 \includegraphics[width=1\linewidth]{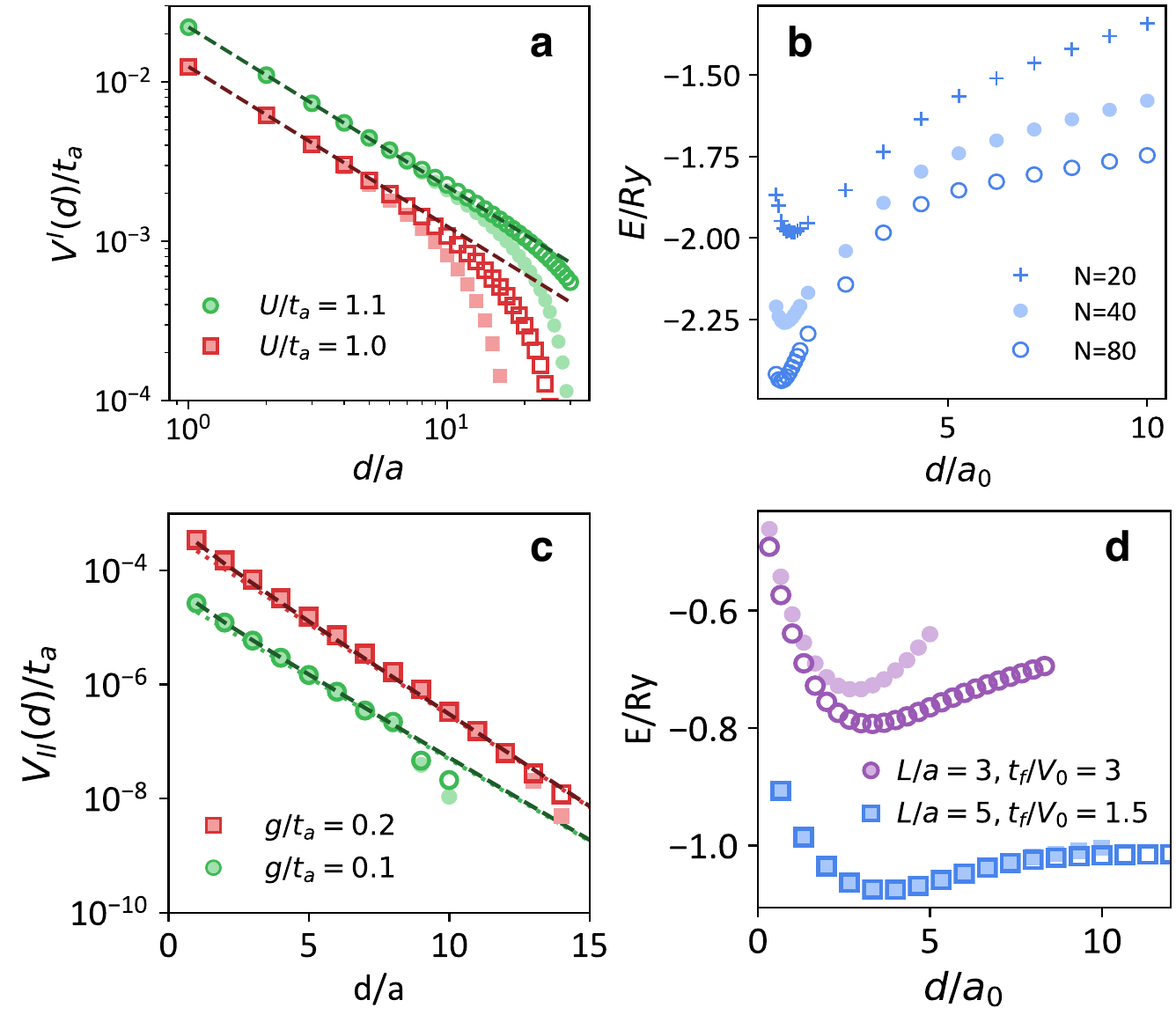}
\caption{ (a) Energy of the single-boson bound state described by the first scheme Eq. \eqref{eq:hamee} as the number of sites $d/a$ separating two fermions is modified. Dashed lines follow the scaling $V\st{I,0}/(d/a)$. (b) Ground state energy of the simulated Hamiltonian for $H_2$ for different lattice sizes and the effective potential $V\st{I}(d)$. (c) Calculation of the repulsion mediated by the second scheme \eqref{eq:2levelHamiltonian} between two fixed fermions separated $d/a$ sites (markers). Dashed lines follows the analytical approximation \eqref{eq:mediatExponential}. Edged markers corresponds to $N=80$ and coloured ones to $N=40$. Here, $U=4.1\, t_a$. (d) Molecular potential for a "pseudomolecule" of hydrogen, where both nuclear attraction and electronic repulsion follow the exponential scaling \eqref{eq:mediatExponential}. Here, edged markers represent $N=60$, coloured ones $N=30$, and $L/a=\pa{2\sqrt{\delta\st{II}/t_a}}^{-1}$. See \cite{Note7} for details.}
\label{fig:H2}
\end{figure}

\emph{Two-fermions model.}
Let us now explore the situation with two fermionic atoms emulating two electrons, where the interelectronic repulsion between them needs to be  mediated. For this, we use an additional bosonic atom trapped in an optical lattice potential with the same geometry as the fermions. First, we start with a simple scheme that only considers one of the bosonic internal states, which allows them to tunnel at a rate $t_a$ to nearest-neighboring sites. As they coexist in the same lattice sites, elastic scattering processes between the bosonic and fermionic atoms occupying the same position induce an on-site repulsion $U$,
\begin{equation}
\label{eq:hamee}
    \ha\st{med,I}=-t_a\,\sum_{\langle \ii, \jj \rangle}a^\dagger_\ii a_ \jj +U\sum_\ii a_\ii^\dagger a_\ii f_\ii^\dagger f_\ii \,,
\end{equation}
that translates into an effective repulsion between the fermions when the effect of the mediating atom is traced-out:
\begin{equation}
\label{eq:hamRep}
   H_{\text{ee}}=\sum_{\ii,\jj}V(\norm{\ii-\jj})f_\ii^\dagger  f_\ii f_\jj^\dagger f_\jj\,,
\end{equation}
To obtain this expression, we assume to be in the regime in which the bosonic atom dynamics is faster than the movement of the fermions. In this first scheme, and for separations $d/a\ll 0.06\, e^{2\pi t_a/U} \ll N$, this effective repulsion corresponds to, $V_{\text{I}}(d)\approx V_{\text{I},0}/(d/a)\,,$ where  $V_{\text{I},0}\approx 6.4 e^{-2\pi t_a/U}t_a$ (see \cite{Note7}).
This simple scheme then mediates an effective repulsion between the two fermionic atoms that scales as $1/r$, matching the dependence of the distance of 3D molecular interactions, but now restricted to 2D~\footnote{Note that this choice of nuclear potential differs from the one encountered in a flatland world, in which Coulomb's law leads to interactions that scale as $\propto \log(r)$.}. We illustrate the dependence of this potential and its effect in the 2D $H_2$ molecule in Figs.~\ref{fig:H2}(a-b), respectively. There, one can observe molecular potentials also for relatively small lattices and assess the error. The continuum limit is obtained in a similar regime than the $H_2^+$ molecule case.

\emph{Many-fermion models:} By increasing the number of fermionic atoms in the lattice while maintaining a single mediating boson, one would see that not all interactions among pairs of fermions are equally weighted, precluding scalability. Intuitively, it is more favourable for the mediating atom to localize among the pair of fermions that are closer to each other, rather than in an equal superposition, so that not all interaction are equally considered. In Ref.~\cite{arguello2019analogue}, this challenge was overcome by including a cavity that symmetrizes these interactions. This cavity interaction is not available in the present, much simplified experimental setup, where interactions are mediated by a hopping atom, instead of a spin-excitation. Another option to induce a pairwise effective repulsion between these fermionic atoms would be Rydberg excitations, that enable for long-range strong atomic interactions. In particular, one can induce dipole-dipole repulsive interactions that depend on their separation as $1/d^3$ for distances smaller than the Rydberg blockade radius ~\cite{Lukin2000,Ravets2014,Saffman2010,Note9}.

Here instead, we present a second scheme that induces pair-wise interactions by including as many mediating bosonic atoms as electrons need to be simulated. This proposal is scalable, at the price of modifying the scaling of the repulsive interaction (see Fig.~\ref{fig:scheme}(d)). For these $N_f$ mediating atoms, we are going to consider two of its long-lived energy levels, that we call $a$ and $b$, separated by an energy shift $\Delta$. Level $b$ experiences an on-site repulsion $U$ when occupying the same site as a fermion, while the atoms in level $a$ live on a shallow lattice that allows them to move with tunneling rate $t_a$. Both levels are coupled through a Raman (or direct) transition of strength $g$. Besides, bosonic atoms in the $b$ level suffer an additional hard-core boson interaction $\abs{W}\gg \abs{U}$ which prevents doubly occupied states. The bosonic Hamiltonian then reads as,
\begin{equation}
\begin{split}
\label{eq:2levelHamiltonian}
\ha\st{med,II}= & -t_b \sum_{\langle \ii , \jj\rangle }b_\ii^\dagger b_\jj -t_a \sum_{\langle \ii , \jj\rangle }a_\ii^\dagger a_\jj +g\sum_\jj (b_\jj^\dagger a_\jj +\hc)\\
&+\Delta \sum_\jj b_\jj^\dagger b_\jj + U\sum_\jj b_\jj^\dagger b_\jj f_\jj^\dagger f_\jj + \frac{W}{2}\sum_\jj b_\jj^\dagger b_\jj^\dagger b_\jj b_\jj 
\end{split}
\end{equation}

In particular, we are interested in the regime in which both levels are weakly coupled $g\ll \Delta $, and when the atomic states trapped in the $a$ lattice hop faster  than in any of the other levels: $t_a \gg t_b \gg t_F$ (see Fig.~\ref{fig:scheme}(b))~\cite{Heinz2019}. This allows one to trace-out the effect of the mediating atoms and write an effective Hamiltonian for the fermions. 
By using as many bosonic atoms as fermions, the hard-core boson interactions leads to a bound state in which all fermionic sites are equally occupied, getting a configuration in which the repulsion among each pair of atoms is equally weighted, as required by Eq.~\eqref{eq:hamee}. For this configuration, the pair-wise mediated interaction scales as,
\begin{equation}
\label{eq:mediatExponential}
    V_{\text{II}}(d)\approx V_{\text{II},0}\, e^{-2d\sqrt{\delta\st{II}}/(a\sqrt{t_a})}\,,
\end{equation}
for $d\sqrt{\delta\st{II}}/(a\sqrt{t_a})\gg1$, where $V_{\text{II},0}\approx \frac{g^4 }{8 \pi t_a^2 \delta\st{II}} \,,$ and $\delta\st{II}=U-4t_a+\ord{g^2/\Delta}$, (see \cite{Note7}).

While this system differs from the molecular Hamiltonian observed in nature, it already captures the key features of the interactions appearing in molecular chemistry: nuclear attraction and electronic repulsion. It is then expected to reveal some of the features of chemical systems, including their electronic correlations. In Fig.~\ref{fig:H2}(c), we show the effective repulsive potential induced by the second scheme for different values of detuning $\delta\st{II}$, that controls the characteristic length of the interaction. In Fig.~\ref{fig:H2}(d), we illustrate the effect that this modified effective repulsion controlled by $\delta\st{II}$ has on two fermionic atoms hopping in the lattice, whose dependence on the distance is also mimicked by the tunable attractive nuclear interaction. This leads to a molecular potential of a "pseudomolecule" of hydrogen, where the bonding length and dissociation limit are observed.

\emph{Conclusions \& Outlook.} To sum up, we have shown how ultra-cold atoms moving in 2D optical lattices can be used to simulate simplified models for quantum chemistry in today’s experimental setups. We have observed that early experiments with a single simulating atom can pursue the timely goal of simulating the simplest discretized atom and molecule in this platform. In richer scenarios, bosonic atoms can mediate an effective repulsion between the simulated electrons, making repulsive interactions more experimentally accessible with state-of-the-art setups. Such simulators open up a number of possibilities for further research. First, they provide an experimental platform for which numerical methods used in quantum chemistry can be adapted and benchmarked. Lessons learnt from these simulators, could then be transferred back into improved algorithms for quantum chemistry. Second, one of the main challenges of these discretized 2D simulators is that their solutions approach the continuum result slower than in the 3D case. Fully characterizing this scaling may well lead to improved protocols that are less sensitive to the system size. Third, while this Letter provides strategies to engineer a pseudochemical Hamiltonian in ultra-cold atoms using bosonic atoms as a mediator, other platforms and strategies may also serve for this purpose. Identifying good candidates to simulate specific interactions in chemistry is a promising open field of research. 

\section*{Acknowledgements}
We acknowledge support from the ERC Advanced Grant QUENOCOBA under the EU Horizon 2020 program (grant agreement 742102). J.A.-L. acknowledges support from 'la Caixa' Foundation (ID 100010434) through the fellowship LCF/BQ/ES18/11670016, the Spanish Ministry of Economy and Competitiveness through the 'Severo Ochoa' program (SEV-2015-0522), Fundaci{\'o} Cellex, Fundaci{\'o} Pere Mir, and Generalitat de Catalunya through the CERCA program. A. G.-T. acknowledges support from the Spanish project PGC2018-094792-B-100 (MCIU/AEI/FEDER, EU) and from the CSIC Research Platform on Quantum Technologies PTI-001. T. S. acknowledges the Thousand-Youth-Talent Program of China and is supported by the NSFC No.11974363. P.Z. acknowledges the EU Quantum Flagship PASQuanS.

\appendix
\renewcommand\thefigure{S\arabic{figure}}  
\setcounter{figure}{0} 
\newpage
\section{Discretization error in 2D}
\label{ap:choosingBohr}
In Fig.~\ref{fig:4Hydrogen2D_ED_atom1} we observed that the discretized solutions of the Hamiltonian approached the analytical result following a scaling $\Delta E \propto \pa{t_f/V_0}^{-1}$. This differs from the three-dimensional case, in which accuracy improves as $\pa{t_f/V_0}^{-2}$ \cite{arguello2019analogue}. To analyze this effect, it is useful to have some insights on how the discretization of the space affects the approach to the continuum solution. A back-of-the-envelope dimensional analysis can be presented for the 2D case, where we consider the ground-state electronic wave-function, $\psi_0(r)=a_0^{-1}\sqrt{2/\pi}e^{-r/a_0}$. 

For the two main sources of discretization error, the calculation of the energy terms is based on integrals that are discretized as a Riemann sum. The difference between this sum and the continuum limit is defined to first order by the second derivative of the integrand. For the Coulomb term, this reads as, 
$$
V_0 \sum_\jj \partial^2_x \pa{|\psi(\rr_\jj)|^2/r}\,,
$$
In the 2D case, this sum does not converge in the continuum limit, and the leading order error corresponds to the diverging term, that is dictated by our choice of the cutoff for the position closest to the nuclei. Normalizing by the Rydberg energy, this error terms scales as $\pa{t_f/V_0}^{-1}$ in 2D, and dominates the scaling of the 2D setup as the effective Bohr radius increases, as numerically observed.

\section{Single-level atom}
\label{ap:effPotential}
\subsection{Single boson localized around one fermion}
As an introductory step to gain intuition, in this section we derive how a mediating boson affects the motion of a single fermion by localizing around it. This is the key ingredient responsible for the effective repulsion appearing when more than one fermion are present, that we derive in the next sections. 
In the limit $t_{F}/t_{B}\ll 1$, one can make an approximation similar to Born-Oppenheimer. For a single fermion occupying the position $\jj_0$, one can then expand the
Hamiltonian $H_{1B}$ in the basis $\left\vert \jj_0\right\rangle _{F}\left\vert
\phi _{\jj_0}\right\rangle _{B}$, where $\left\vert \jj_0\right\rangle
_{F}=f_{\jj_0}^{\dagger }\left\vert 0\right\rangle _{F}$ and $\left\vert
\phi _{\jj_0}\right\rangle _{B}$ is the ground state of $_{F}\left\langle
\jj\right\vert (H_{1B})\left\vert \jj\right\rangle _{F}$ where, in the continuum limit, $H_{1B}$ takes the form, $\sum_\kk \omega_\kk a_\kk^\dagger a_\kk +U a_\jj^\dagger a_\jj f_\jj ^\dagger f_\jj$, being $\omega(\kk)=-2t_b \pa{\cos k_x + \cos k_y}$ the dispersion relation for a free boson.

In the single fermion subspace, let us start choosing the fermion to be positioned in $\jj_0=\mathbf{0}=(0,0)$. The eigenstate writes as $\beta _{\lambda
}^{\dagger }f_{\mathbf{0}}^{\dagger }\left\vert
0\right\rangle $, where the bosonic operator $\beta _{\lambda }^{\dagger
}=\sum_{\kk}\phi _{\lambda }(\kk)a_{\kk}^{\dagger }$. The Schr\"{o}dinger equation
writes as, 
\begin{equation}
\omega_\kk \, \phi (\kk)+U\, \phi(\mathbf{0})=E_B\, \phi (\kk),
\label{SE1}
\end{equation}%
where $\phi(\jj)=1/N\sum_\kk e^{-i\kk\jj} \phi(\kk)$.

In general, for the bound state,%
\begin{equation*}
\left\vert \phi _{\jj_0}\right\rangle _{B}=\sum_{\ii}\phi
_{\jj_0}(\ii)a_{\ii}^{\dagger }\left\vert 0\right\rangle _{B}\,,
\end{equation*}%
describes the single boson localized around the fermion and its bound state energy $E_{B}$ is determined by%
\begin{equation}
\label{eq:1boundState}
U^{-1}=\frac{1}{N}\sum_{k}\frac{1}{E_{B}-\omega_{\kk}}\,.
\end{equation}%

\begin{figure}[tbp]
\centering
 \includegraphics[width=0.9\linewidth]{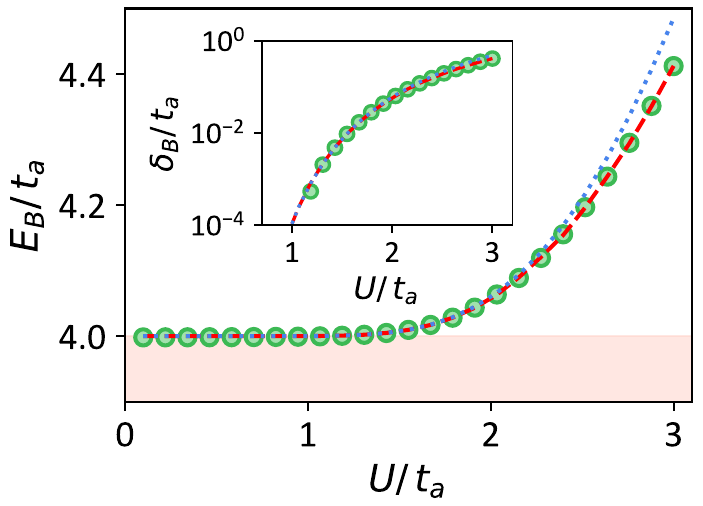}
\caption{Single-fermion bound state energy $(E_B)$ as a function of the fermion-boson interaction $U$, as compared to the solution dictated by \eqref{eq:1boundState}. Markers represent the ED calculation, while $\Sigma(z,0)$ is evaluated using the analytical solution \eqref{eq:sigzAn} (red dashed line) and the approximation \eqref{eq:sigzApr} (blue dotted line). Inset shows the energy separation to the band edge, $\delta_B=E_B-4t_a$. Here, $N=100$ and finite-size effects appear for $U/t_a\lesssim 1$.}
\label{fig:boundStEmas}
\end{figure}
Its wavefunction writes as,%
\begin{equation}
\label{eq:apWave}
\phi _{\jj_{0}}(\jj)=\frac{1}{\sqrt{\mathcal{N}_{1B}}}\frac{1}{N}\sum_{\kk}\frac{%
e^{-i\kk(\jj-\jj_{0})}}{E_{B}-\omega_{\kk}}\,,
\end{equation}%
where the normalization factor,%
\begin{equation}
\mathcal{N}_{1B}=\frac{1}{N}\sum_{\kk}\frac{1}{(E_{B}-\omega_{\kk})^{2}}.
\end{equation}

We define a pair creation operator $F_{\jj}^{\dagger }$, which generates the
local Wannier mode $F_{\jj}^{\dagger }\left\vert 0\right\rangle =\left\vert
j\right\rangle _{F}\left\vert \phi _{\jj}\right\rangle _{B}$ by acting on the
vacuum state. In terms of $F_{\jj}$ and $F_{\jj}^{\dagger }$, the Hamiltonian
under this approximation becomes,%
\begin{equation}
H_{\mathrm{BO}}=\sum_{\jj}[E_{B}F_{\jj}^{\dagger }F_{\jj}-\tilde{t}%
_{F}(F_{\jj}^{\dagger }F_{\jj+1}+\mathrm{H.c.})]\,,
\end{equation}%
by projecting on the bound state energy surface, where the effective hopping
strength,
\begin{equation}
\tilde{t}_{F}=t_{F}\left\langle \phi _{\jj}\left\vert \phi _{\jj+1}\right\rangle
\right. =\frac{t_{F}}{\mathcal{N}_{1B}}\frac{1}{N}\sum_{\kk}\frac{e^{-ik_x}}{%
(E_{B}-\omega_{\kk})^{2}}\,,
\end{equation}%
of the bound boson-fermion pair is determined by the Franck-Condon
coefficient $\left\langle \phi _{\jj}\left\vert \phi _{\jj+1}\right\rangle
\right. $, i.e., the overlap of the bosonic Wannier states. 

\subsection{Single boson localized around two fermions}
By introducing a second fermion, the boson forms a bound-state whose energy depends on this interfermionic separation, inducing an effective repulsion between these two fermions. Aided by the intuition gained in the previous section, here we characterize the properties of this bosonic bound-state.

In the single-boson subspace, the eigenstate writes as $\beta _{\lambda
}^{\dagger }f_{\jj_{1}}^{\dagger }f_{\jj_{2}}^{\dagger }\left\vert
0\right\rangle $, where the bosonic operator $\beta _{\lambda }^{\dagger
}=\sum_{\kk}\phi _{\lambda }(k)a_{\kk}^{\dagger }$. The Schr\"{o}dinger equation
leads to%
\begin{equation}
\omega_{\kk}\phi _{\lambda
}(\kk)+C_{1}e^{-i\kk \jj_{1}}+C_{2}e^{-i\kk \jj_{2}}=E_{\lambda }\phi _{\lambda }(\kk ),
\label{SE}
\end{equation}%
with parameters,%
\begin{eqnarray}
C_{1} &=&\frac{U}{N}\sum_{\kk}e^{i\kk \jj_{1}}\phi _{\lambda }(\kk),  \notag \\
C_{2} &=&\frac{U}{N}\sum_{\kk}e^{i\kk \jj_{2}}\phi _{\lambda }(\kk).
\end{eqnarray}
The bound state solution%
\begin{equation}
\phi _{\pm }(k)=\frac{C_{1}e^{-i\kk \jj_{1}}+C_{2}e^{-i\kk \jj_{2}}}{E_{\pm
}-\omega_{\kk}}\,,
\end{equation}%
of Eq. (\ref{SE}) gives rise to the self-consistent equation%
\begin{eqnarray}
C_{1} &=&\frac{U}{N}\sum_{\kk}\frac{C_{1}+C_{2}e^{i\kk\dd}}{E_{\pm
}-\omega_{\kk}},  \notag \\
C_{2} &=&\frac{U}{N}\sum_{\kk}\frac{C_{1}e^{i\kk\dd}+C_{2}}{E_{\pm
}-\omega_{\kk}},
\end{eqnarray}%
which determines the relation $C_{1}=\pm C_{2}$. Focusing on the bound state on the upper-band, that provides the repulsive interaction, and defining $\tilde k_{x,y}\equiv -\pi + k_{x,y}$, the bound state energy $E\st{up}$ corresponds to,
\begin{equation}
\label{eq:2boundState}
U^{-1}=\frac{1}{N}\sum_{\kk}\frac{1 + e^{i\kk\dd}}{E\st{up}-\omega_{\tilde \kk}}.
\end{equation}%
This equation encodes how the energy of the bound state depends on the interfermionic separation. Note that $\dd$ is a 2D-vector with integer components.

Equating \eqref{eq:1boundState} and \eqref{eq:2boundState}, one gets,
\begin{equation}
\frac{1}{N}\sum_{\kk}\frac{1 }{E_{B}-\omega_{\kk}}=\frac{1}{N}\sum_{\kk}\frac{1 + e^{i\kk\dd}}{E\st{up}-\omega_{\tilde\kk}}.
\end{equation}%
 The solution to this equation admits a solution given by a recurrence relation on $\dd$~\cite{Katsura1971}. Using instead the expansions derived in Sec.~\ref{sec:apFirtInt} and~\ref{sec:apSecInt}, one gets for $d/a\ll 1/\sqrt{\delta_B/t_a}$),
\begin{equation}
\label{eq:deltaMasd}
    \delta\st{up} = E_+-4t_a\approx \frac{2\sqrt{\delta_B}}{d}e^{-\gamma}\,,
\end{equation}
where $\delta_B=E_B-4t_a\approx 2^5 e^{-4\pi t_a/U} t_a$ 
, and $\gamma \approx 0.577\ldots$ is the Euler-Mascheroni constant. 

This simple model then provides an effective repulsion between the two fermions that scales as $\delta\st{up}(d)/t_a \propto V\st{0,I}/d $ with  $V\st{0,I}=2^{7/2}e^{-\gamma-2\pi t_a/U}t_a$. 

\begin{figure}[tbp]
\centering
 \includegraphics[width=0.95\linewidth]{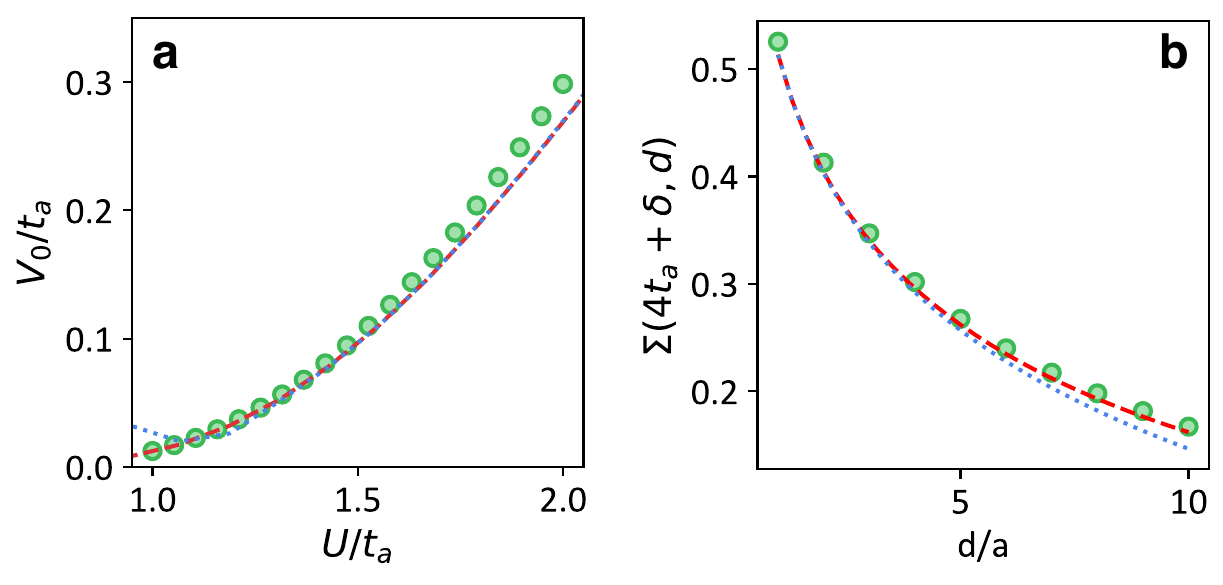}
\caption{(a) This initial value $V_0$ is calculated for different choices of bosonic-fermionic interaction $U$, and compared to the approximation \eqref{eq:deltaMasd}. In this equation, the value of $\delta_B$ is calculated using the exact solution \eqref{eq:sigzAn} (red dashed line), and using ED (blue dotted line). For the numerical calculation, $N=100$. (b) Numerical evaluation of \eqref{eq:2dint} for $N=100$ (markers), as compared to the analytical solution in the continuum \eqref{eq:2dintInt}(red dashed line) and approximation \eqref{eq:2dintAprox} (blue dotted line). Here, $\delta/t_a=0.002$.}
\label{fig:vd}
\end{figure}

From the wavefunction \eqref{eq:apWave} and the expansion in Sec. \ref{sec:apSecInt}  one sees that the characteristic length of the bound states is $L\st{I}/a\approx \pa{\delta_B/t_a}^{-1/2}$. For the previous expansions in \eqref{eq:deltaMasd} to be valid, one needs to satisfy the regime $d/a\ll L\st{I}/a$. To prevent finite size effects, it is also necessary, that $L\st{I}/a\ll N$. To illustrate this, in Fig. \ref{fig:boundStEmas} we observe that this expansion for $\delta_B/t_a$ is valid for $U/t_a >1$, so that $L\st{I}/a\ll N=100$. In Fig. \ref{fig:vd} we also confirm that for this size, the scaling $1/d$ is maintained for $d/a\ll 10$, so that $d/a  \ll L\st{I}/a$.

One can now see that this pairwise interaction does not maintain when more than two fermions are present. To reach this scalability, in Appendix \ref{ap:2levelNboson} we will consider a second internal level of the mediating atom.

\subsection{Calculation of the first integral in \eqref{eq:2boundState}}
\label{sec:apFirtInt}
Defining the energy and length units $t_a\equiv1$, $a\equiv 1$ in the coming sections, let us now calculate, $$   \Sigma(z,0) =\frac{1}{N^2} \sum_\kk  \frac{1}{z-\omega_\kk}\,. $$ 
One can write an analytical solution \cite{Katsura1971},
\begin{equation}
\label{eq:sigzAn}
\Sigma(z,0) = 2\,K\co{4/z}/(\pi z)\,,
\end{equation}
where $K[m]=\int_0^{\pi/2} d\theta \pa{1-m^2\sin^2(\theta)}^{-1/2}$ is the complete elliptic integral of the first kind for $\abs{m}\leq 1$~\cite{abramowitz1972integration}. For values $z=4+\delta$ close to the band-gap $(\delta>0$ and $\abs{\delta}\ll 1$), one can define,
\begin{equation}
    \label{eq:sigzApr}
    \Sigma(z,0) \approx (5 \log 2 - \log \delta)/(4 \pi) + \ord{\delta ^2}\,.
\end{equation}

\subsection{Calculation of the second integral in \eqref{eq:2boundState}}
\label{sec:apSecInt}
In order to extract the scaling of \eqref{eq:2boundState} for frequencies close to the band-gap, it is useful to explore the continuous version of this sum. This will introduce a divergence, that was prevented by the natural cutoff of the lattice. 

Now, we are interested in the calculation of,
\begin{equation}
\label{eq:2dint}
\Sigma(z,\dd)=\frac{1}{N^2}\sum_{\kk} \frac{e^{i\kk\dd}}{z-\omega(\kk)}\,,
\end{equation}
for $D=\co{0,2\pi}^{\otimes 2}$.

In the limit $\kk\dd \gg1$, we can expand the dispersion relation for frequencies close to the upper band-edge, [$(k_x,k_y)=(\pi,\pi)$]. Taking the translation $\tilde k_{x,y}\equiv -\pi + k_{x,y}$, we expand $\omega(\tilde \kk)\approx 4-\tilde \kk^2$, and extend the integration domain to infinite. Note that the numerator $e^{i\kk \dd}$ prevents the otherwise divergent integral, and the frequency shift introduces a sign factor, $e^{i \bm{\pi}\dd}$, that does not enter in the mediated potentials for the strategies presented in this Letter. W.l.o.g., we align vector $\rr$ in the $z$-axis, and use spherical units,
\begin{equation}
\label{eq:2dintInt}
\begin{split}
\Sigma(z,\dd)
 = e^{i \bm{\pi}\dd} K_0\co{d\sqrt{z-4}}/( 2\pi )\,,
\end{split}
\end{equation}
where $K_n[x]$ is the modified Bessel function of the second kind~\cite{abramowitz1972integration} and $d\equiv \norm{\dd}$. For small arguments ($0<x\ll1$), 

\begin{equation}
    \label{eq:2dintAprox}
    K_0[x]\approx -\log(x/2)-\gamma\,.
\end{equation}

\section{Mediating atoms with two long-lived states}
\label{ap:2levelNboson}
When more than two fermionic atoms are introduced, the effective repulsion mediated in the previous section by the single-boson bound-state is not purely described by the pair-wise separation between each pair of fermions. To gain this feature, let us introduce in this Section a modified scheme, where we consider two internal levels of as many mediating atoms as fermions there are in the system. We will denote the two levels as $b$ and $a$. Atoms in $b$  level experience an on-site repulsion when occupying the same site of a fermion, while atoms in state $a$ live on a shallow lattice that allows them to hop with tunneling rate $t_a$. Both levels are coupled through a Raman transition of strength $g$ and are shifted by energy $\Delta$. In order to equally account for repulsion among each pair of fermionic atoms, we include an on-site repulsion $W$ among them when they occupy the same lattice site, obtaining the mediating Hamiltonian,
\begin{equation}
\begin{split}
\label{eq:hamMultipleBosons}
H&\st{med,II} =\Delta \sum_\jj a_\jj^\dagger a_\jj-t_b \sum_{\langle i , j\rangle } b_\ii^\dagger b_\jj -t_a \sum_{\langle \ii , \jj\rangle }a_\ii^\dagger a_\jj\\ 
&+ U \sum_\jj b_\jj^\dagger b_\jj  f_\jj^\dagger f_\jj+g\sum_\ii (b_\jj^\dagger a_\jj +\hc) +\frac{W}{2} \sum_\jj b_\jj^\dagger b_\jj^\dagger  b_\jj b_\jj \,.
\end{split}
\end{equation}

Intuitively, mediating atoms localize around the fermionic positions, and double occupations are prevented by the hard-core boson interaction $W\gg U$. This then creates a bound-state in which each mediating atom localizes in a different fermionic position. As compared to the previous scheme, hopping from one fermion to the others now becomes a fourth-order process in the coupling $g$ between the two atomic metastable states, as the movement of two mediating atoms is needed. 

In particular, we are interested in the regime in which both levels are weakly coupled $g/\Delta \ll 1$, and atoms in level $a$ hop in a lattice much more shallow than the rest: $t_f\ll t_b \ll t_a$. As it occurred in the previous case, this last inequality allows to trace-out the effect of the mediating atom, writing an effective Hamiltonian for the fermions, $\sum_{\ii\jj} V(|\ii-\jj|)f_\ii^\dagger  f_\ii f_\jj^\dagger f_\jj$. Let us now derive this regime using perturbation theory for $g/\Delta \ll 1$ and $N_f$ fermions occupying fixed positions $\jj_1 \ldots \jj_{N_f}$. For this, let us separate the bosonic Hamiltonian \eqref{eq:hamMultipleBosons}, as  $ H_{BN}= H_0 + H_I$, where
\begin{equation}
\label{eq:pertHamilt}  
\begin{split}     
H_0=&\Delta \sum_\jj a_\jj^\dagger a_\jj-t_b \sum_{\langle i , j\rangle } b_\ii^\dagger b_\jj -t_a \sum_{\langle \ii , \jj\rangle }a_\ii^\dagger a_\jj\\ 
&+ U \sum_\jj b_\jj^\dagger b_\jj  f_\jj^\dagger f_\jj+\frac{W}{2} \sum_\jj b_\jj^\dagger b_\jj^\dagger  b_\jj b_\jj \,, \\     
H_I = & g\sum_\ii (b_\jj^\dagger a_\jj +\hc)\,.
\end{split}
\end{equation}
In particular, we are interested in the energy correction of the bound-state  $ \ket{\psi_{B,\text{II}}} = \prod_{i=1}^{N_f} b^\dagger_{\jj_i} \ket{0}$, that depends on the interfermionic positions. For this, we need to expand the perturbed Hamiltonian. One can see that only even orders enter the calculation, and expanding to fourth order,
\begin{equation}
\label{eq:expansionHam}
\begin{split}
       E_{B,\text{II}}& \ket{\psi_{B,\text{II}}}  = \Big( H_0+H_I\frac{1}{E-H_0}H_I\\
        &+H_I\frac{1}{E-H_0}H_I\frac{1}{E-H_0}H_I\frac{1}{E-H_0}H_I \Big) \ket{\psi_{B,\text{II}}}\,,
\end{split}
\end{equation}
one gets the equation,
\begin{equation}
\label{eq:expansion}
\begin{split}
    &E_{B,\text{II}}  =N_f U + N_f\frac{g^2}{N^2}\sum_\kk \frac{1}{E_{B,\text{II}}/N_f-\Delta-\omega_\kk} \\
    &+ \frac{2g^4}{N^4} \sum_{i\neq j=1}^{N_f}\sum_{\kk,\qq} \frac{1+e^{i(\kk-\qq)(\rr_i-\rr_j)}}
    {\pa{E_{B,\text{II}}/N_f-\Delta-\omega_\kk}^2 \pa{E_{B,\text{II}}/N_f-\Delta-\omega_\qq}
    }\,.
\end{split}
\end{equation}

This latter term originates from the pairwise repulsion introduced by the fourth-order correction of two mediating atoms swapping the fermionic position they localize around. This then leads to an effective pairwise potential, $\sum_{i\neq j=1}^{N_f}V_{\text{II}}(\norm{\rr_i-\rr_j)})f_\ii^\dagger  f_\ii f_\jj^\dagger f_\jj$, where
\begin{equation}
\label{eq:pretwo-levelDistance}
\begin{split}
V_{\text{II}}(\dd)\approx &\frac{2g^4}{N^4} 
    \pa{ \sum_{\kk}\frac{e^{i\kk\dd}}
    {\pa{E_{B,\text{II}}/N_f-\Delta-\omega_\kk}^2 }}\\
    &\times
    \pa{\sum_{\qq}  \frac{e^{-i\qq\dd}}
    {E_{B,\text{II}}/N_f-\Delta-\omega_\qq}}\,.
    \end{split}
\end{equation}

These two independent sums can be calculated as in Sec.~\ref{sec:apSecInt}. Note that the alternating sign derived in Sec.~\ref{sec:apSecInt} cancels after the double product $e^{i\kk\dd}e^{-i\qq\dd}$. Using that $\partial_x K_0[x]=-K_1[x]$, one obtains,
$
V_{\text{II}}(d)\approx \frac{2g^4}{(2\pi)^2} K_0\co{d\sqrt{\delta\st{II}}}\frac{d}{2\sqrt{\delta\st{II}}}	K_1\co{d\sqrt{\delta\st{II}}}\,,$
which, to lowest order in the regime $d\sqrt{\delta\st{II}}>1$, scales as,
\begin{equation}
\label{eq:two-levelDistance}
V_{\text{II}}(d) \approx  \frac{g^4}{8\pi\delta\st{II}}   e^{-2d\sqrt{\delta\st{II}}}\,.
\end{equation}
This then leads to a pairwise repulsion between the fermionic atoms that decays exponentially with their separation, following a decay length $L\st{II}\equiv \pa{2\sqrt{\delta\st{II}}}^{-1/2}$. In Fig. \ref{fig:H2}(c), we approximate $\delta\st{II}$ to second order as \begin{equation}
\label{eq:delta2}
    \delta\st{II}\approx \delta + E_{B,\text{II}}^{(2)}(\delta)/N_f\,,
\end{equation}
where $E_{B,\text{II}}^{(2)}(\delta)$ approximates the second order correction in \eqref{eq:expansion} as,
\begin{equation}
\label{eq:2levelFirstOrderExpansion}
     E_{B,\text{II}}^{(2)}(\delta)=N_f\frac{g^2}{N^2}\sum_\kk \frac{1}{U-\Delta-\omega_\kk}\,,
\end{equation}
that can be expanded as in \eqref{eq:sigzApr}.

\section{Numerical methods}
\subsection{Exact diagonalization}
\label{ap:choosingDist}
Once the kinetic term is approximated as a nearest-neighbor hopping term \eqref{eq:hamKin}, the Hamiltonian can be conveniently written in a position basis and the ground-state obtained using exact diagonalization (ED).

In Fig. \ref{fig:4Hydrogen2D_ED_atom1} we use this approach to calculate the energies $\lambda_n$ associated to the lowest part of the spectrum of Hamiltonian \eqref{eq:Hsimul} for different choices of the ratio $t_f/V_0$. These energies are shifted to correct the shift induced by the nearest-neighbor approximation, $\omega(\kk)\approx 1-k^2/2$. The result is divided by the Rydberg energy; this is, $E_n/\text{\Ry}=\co{\lambda_n+N_f\cdot (2t_f)}/(V_0^2/t_f)$ where, in this case $N_f=1$. The same strategy is applied to calculate the fermionic potential in Fig. \ref{fig:Hmas}(a), where only one mediating atom is involved.

This approach is also used in Fig. \ref{fig:Hmas}(a) to calculate the ground-state energy of an Hydrogen cation for a given internuclear separation $d/a_0$. In addition to the previous shift, nuclear repulsion $V_0/(d/a)$ needs to be included before expressing the result in Rydberg energies. Similarly to the atomic case, for a fixed interatomic distance, accuracy improves by increasing the effective Bohr radius $a_0/a=t_F/V_0$, up to the point in which finite-size effects become relevant. The number of lattice sites separating the nuclear positions $d/a$ is then adjusted accordingly, identifying the optimal separation value as the one giving the lowest ground-state energy (see Fig. \ref{fig:Hmas}(b)).

The same strategy is also applied to obtain the ground state energy of $H_2$ in Fig. \ref{fig:H2}(b). The main difference is that now $N_f=2$, and further simplifications can be made taking into consideration the fermionic statistics. As each fermion can occupy $N^2$ sites which, together with the fermionics statistic $\{f_\ii,f_\jj^\dagger\}=\delta_{\ii,\jj}$ leads to a Hilbert space of space of size $N^2(N^2-1)/2$. 

ED is also used in \ref{fig:H2}(b) to calculate the effective potential mediated by a single bosons, for fixed fermionic positions separated by $d/a$ sites and centered in the lattice.

The exponential decaying potential explored in Fig. \ref{fig:H2}(d), requires a more careful analysis, as the natural rescalings to the Bohr radius and Rydberg energy does not apply now. In particular, three parameters can be independently tuned: the fermionic hopping $t_f$, the interacting potential $V_0$ and the decay length $L\st{II}/a$. As compared to the previous case, one can remap $V_0\to V_0^2/t_f$ and $L\st{II}\to L\st{II}t_f/V_0$, so that the final result is still dimensionless when normalizing the energies by the previous definition of Bohr radius $V_0^2/t_f$. As an illustration, in this Figure \ref{fig:H2}(d), $L\st{II}/a=5$ is chosen, and $t_F/V_0$ is fixed as the ratio providing maximum accuracy for the atomic case (one fermion an one nuclei) hopping in a lattice of side $\fl{N/2}$, so that the dissociation limit is properly captured. Modifying the separation $d/a$ between nuclear positions then allows to scan the different internuclear separations $d/a_0=(d/a) \cdot (V_0/t_f)$ for this fixed value of $t_F/V_0$.

\subsection{Imaginary time evolution}
\label{ap:IT}
For the calculation of the effective potential mediated by the two metastable levels of atoms in \ref{fig:H2}(c), we use Imaginary Time Evolution (ITE). This is a useful strategy to numerically obtain the ground state of a gapped Hamiltonian with purely positive eigenvalues, and consists on iteratively evolving an initially random state as, $e^{-H\cdot t}$. After each iteration the resulting state is normalized, and the contribution of the excited states is mostly reduced.

 In more detail, one of the advantages of this method is that rather than writing the entire evolution operator $\mathcal{O}(N^4)$, one can choose to work in a diagonal basis, so that only $\mathcal{O}(N^2)$ terms are needed to describe the state at each point in time. From the computational perspective, this is specially useful when facing the multielectronic case. In principle, to calculate the interaction among $N_b$-bosons one would need a state with $(N^2)^{N_b}$ entries. Using ED, one would need to write the Hamiltonian, of size $(N^2)^{N_b} \times (N^2)^{N_b}$. In contrast, evolving the state in imaginary time evolution only needs to store the diagonal terms [with size $(N^2)^{N_b}$], once the state is expressed in a basis that commutes with the terms of the Hamiltonian. For our particular case, this corresponds to the position representation for the on-site interactions, and momentum representation for the kinetic term. The Hamiltonian $H_{\text{nuc}}$ is already diagonal in position basis, and one can define a momentum basis,
\begin{equation}
    f_\kk^\dagger (b_\kk^\dagger)=\frac{1}{N} \sum_{\jj} e^{-i\kk\jj} f_\jj^\dagger (b_\jj^\dagger)\,,
\end{equation}
where $H_K$ reads as $H_K=\sum_\kk \omega_{\kk,f} f_\kk^\dagger f_\kk$, being $\omega_{\kk,f}=-2t_F\pa{\cos(k_x)+\cos(k_y)}$ the dispersion relation. This induces a periodic boundary condition in the lattice, which does not affect the calculation as long as finite-size effects are prevented. To confirm that is the case, for each choice of parameters we check that the same result is obtained for the single-boson case using ED, evidencing that boundary conditions are not affecting the result.

To calculate the ITE of Hamiltonian \eqref{eq:2levelHamiltonian}, a constant energy shift is added to $H$ during the calculation to make all the spectrum positive, which is later subtracted at the end of the calculation. To evaluate the operation, $\psi(t)=e^{-Ht}\psi(0)$ we use a Suzuki-Trotter~\cite{Schmid2006} expansion of the first kind, dividing the evolution in $n$ steps as $e^{-Ht}\approx \prod_{k=1}^{n-1} e^{-H\Delta t}+\ord{\Delta t}$, and $t_k=k\cdot \Delta t/t$. For each of these steps, we calculate
\begin{equation}
\begin{split}
\label{eq:ITiteration}
  e^{-H\Delta t}\psi(t_k) \approx & \text{IFFT} \co{e^{-H_K \Delta t}\, \text{FFT} \pa{e^{-H_{\text{R}}\Delta t} e^{-H_{g}\Delta t} \psi(t_{k-1}}}\\ 
  & +\ord{\Delta t^2}\,,
  \end{split}
\end{equation}
where (I)FFT indicates the (Inverse) Fast Fourier Transformation, and normalize the resulting state. Here, $H_{\text{R}}$ denote the terms that are diagonal in the position basis, and $H_{\text{K}}$ the ones in momentum basis. $H_g$ denotes the coupling term, whose exponential can be directly calculated noting that, $e^{-g(a_\jj^\dagger b_\jj+\hc)\Delta t}=\cosh(g\Delta t)(a_\jj^\dagger a_\jj+b_\jj^\dagger b_\jj)-\sinh(g\Delta t)(a_\jj^\dagger b_\jj+\hc)$. We iterate this procedure until the overlap between $\psi(t_{k-1})$ and $\psi(t_{k})$ is smaller than $10^{-5}$. We initialize the algorithm with a random state for the smallest value of $t_F/V_0$, and use this converged solution as the initial state for the next configuration of $t_F/V_0$.

In our second scheme, $N_f$ atoms with two long-lived states are used to mediate the interaction among $N_f$ fermions. For a given fermionic configuration, we desire to numerically calculate the bound state, and compare it to the analytical expansion previously introduced in Eq. \eqref{eq:pretwo-levelDistance}. For this calculation, we use the ITE method (Sec. \ref{ap:IT}), where now, each of the $N_f$ mediating atoms can occupy any of the 2 levels at any of the $N^2$ lattice sites, which \emph{a priori} accounts for states of size $(2N^2)^{N_f}$. To reduce this space, we assume that $\abs{g/(U-\Delta)} \ll1$, so that level $b$ is only populated in the sites where they interact with the fermions, and $\abs{W/U}\gg 1$, so that two mediating atoms in level $b$ do not coexist in the same lattice site. For a configuration of 2 (3) fermions in sites $\rr,\sss(,\ttt)$, and given the indistinguishability of the mediating atoms, we can further reduce the Hilbert space to states written in the basis collected in Tables I and II. 

Within this basis, in Fig. \ref{fig:H2} we calculate how the energy of the bosonic ground-state energy $E(d)$ depends on their separation $d$ between two fermionic atoms fixed in lattice sites $[N/2-d/2,N/2]$ and $[N/2+d/2,N/2]$, following the same strategy used in the previous case for $N_f=2$.

\begin{figure}[tbp]
\centering
  \includegraphics[width=0.50\linewidth]{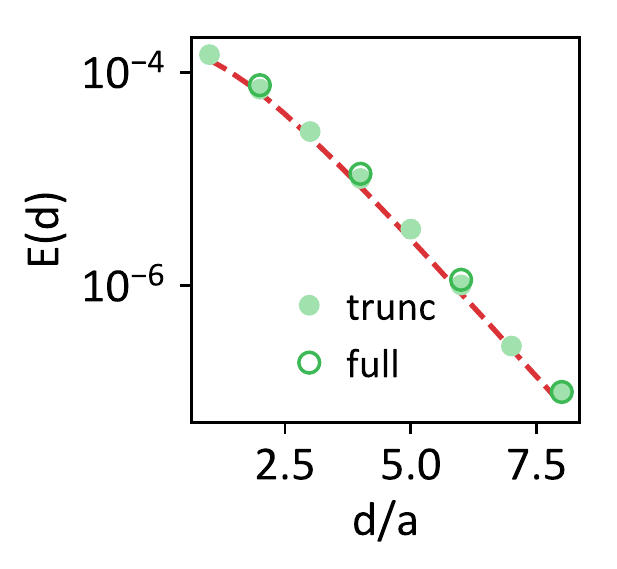}
   \includegraphics[width=0.40\linewidth]{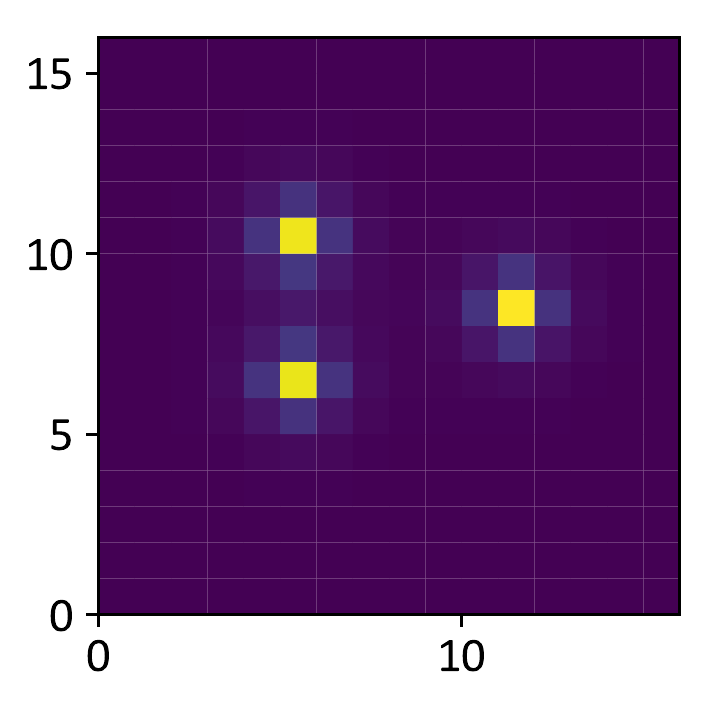}
\caption{(a) Bound-state energy dependence for 3 bosons affected by 3 fermionic atoms occupying fixed positions describing an isosceles triangle of basis 4 sites, and variable height. Contour markers refer to the full basis in which 3 bosonic atoms can simultaneously occupy level $a$, and full markers refer to the truncated basis. Dashed line follows the solution in Eq. \eqref{eq:pretwo-levelDistance}. (b) Occupation of state $a$ for a height of 6 sites. Parameters: $N=16,\, \delta/t_a=g/t_a=0.3t_a$.}
\label{fig:3bosons}
\end{figure}

In the case $N_f=3$, we observe that the biggest demand on computational memory corresponds to describing processes in which the three mediating atoms simultaneously populate the $a$-level. Such processes scale as $[g/(U-\delta)]^6$ in perturbation theory, and are subleading when compared to the second-order terms. Therefore, truncating $0\leq N_a \leq 2$ would allow to push the calculation at a marginal error (see Table II). 

To confirm this intuition, in Fig. \ref{fig:3bosons}(a) we use ITE to calculate the bosonic bound state for 3 fermions describing a triangular isosceles configuration. For moderate sizes ($N=16$), we compare the numerical result given by this truncated space to the one obtained for the total basis. As desired, we observe that (1) the truncation to the space with up to 2 excitations in state $a$ does not modify the solution, and (2) the scaling is in agreement with the calculation for a pairwise repulsion given by \eqref{eq:two-levelDistance}.

\begin{table}[]
\label{tab:basisChoice2}
\begin{tabular}{cccc}
\hline
$N_b$              & $N_a$              & State                                           & Size                    \\ \hline
2                  & 0                  & $b^\dagger_r b^\dagger_s |0\rangle$ & 1                       \\ \hline
\multirow{2}{*}{1} & \multirow{2}{*}{1} & $b^\dagger_r a^\dagger_m |0\rangle$ & \multirow{2}{*}{$2N^2$} \\
                   &                    & $b^\dagger_s a^\dagger_m |0\rangle$ &                         \\ \hline
0                  & 2                  & $a^\dagger_m a^\dagger_n |0\rangle$ & $N^2(N^2+1)/2$               \\ \hline
\end{tabular}
\caption{\textbf{Two mediating atoms.} Basis used to describe states in which $N_b$ of the two mediating atoms occupy level $b$ in the fermionic sites $\rr,\sss$, and $N_a$ atoms are in level $a$ for any choice of sites $\mm,\nn$, in the lattice of size $N\times N$.}
\end{table}

\begin{table}[]
\label{tab:basisChoice3}
\begin{tabular}{cccc}
\hline
$N_b$              & $N_a$              & State                                           & Size                       \\ \hline
3                  & 0                  & $b^\dagger_r b^\dagger_s b^\dagger_t |0\rangle$ & 1                          \\ \hline
\multirow{3}{*}{2} & \multirow{3}{*}{1} & $b^\dagger_r b^\dagger_s a^\dagger_m |0\rangle$ & \multirow{3}{*}{$3N^2$}     \\
                   &                    & $b^\dagger_s b^\dagger_t a^\dagger_m |0\rangle$ &                            \\
                   &                    & $b^\dagger_t b^\dagger_r a^\dagger_m |0\rangle$ &                            \\ \hline
\multirow{3}{*}{1} & \multirow{3}{*}{2} & $b^\dagger_r a^\dagger_m a^\dagger_n |0\rangle$ & \multirow{3}{*}{$3N^2(N^2+1)/2$} \\ 
                   &                    & $b^\dagger_s a^\dagger_m a^\dagger_n |0\rangle$ &                            \\
                   &                    & $b^\dagger_t a^\dagger_m a^\dagger_n |0\rangle$ &                            \\ \hline
0                  & 3                  & $a^\dagger_m a^\dagger_n a^\dagger_p |0\rangle$ & $(N^2+2)(N^2+1)N^2/6$                  \\ \hline
\end{tabular}
\caption{\textbf{Three mediating atoms.} Similarly to Table I, here we define the basis associated to $N_b$ of the three mediating atoms occupying level $b$ in the fermionic sites $\rr,\sss,\ttt$, and $N_a$ atoms being in level $a$ for any choice of sites $\mm,\nn,\pp$, in the lattice of size $N\times N$.}
\end{table}


\begin{thebibliography}{43}%
\makeatletter
\providecommand \@ifxundefined [1]{%
 \@ifx{#1\undefined}
}%
\providecommand \@ifnum [1]{%
 \ifnum #1\expandafter \@firstoftwo
 \else \expandafter \@secondoftwo
 \fi
}%
\providecommand \@ifx [1]{%
 \ifx #1\expandafter \@firstoftwo
 \else \expandafter \@secondoftwo
 \fi
}%
\providecommand \natexlab [1]{#1}%
\providecommand \enquote  [1]{``#1''}%
\providecommand \bibnamefont  [1]{#1}%
\providecommand \bibfnamefont [1]{#1}%
\providecommand \citenamefont [1]{#1}%
\providecommand \href@noop [0]{\@secondoftwo}%
\providecommand \href [0]{\begingroup \@sanitize@url \@href}%
\providecommand \@href[1]{\@@startlink{#1}\@@href}%
\providecommand \@@href[1]{\endgroup#1\@@endlink}%
\providecommand \@sanitize@url [0]{\catcode `\\12\catcode `\$12\catcode
  `\&12\catcode `\#12\catcode `\^12\catcode `\_12\catcode `\%12\relax}%
\providecommand \@@startlink[1]{}%
\providecommand \@@endlink[0]{}%
\providecommand \url  [0]{\begingroup\@sanitize@url \@url }%
\providecommand \@url [1]{\endgroup\@href {#1}{\urlprefix }}%
\providecommand \urlprefix  [0]{URL }%
\providecommand \Eprint [0]{\href }%
\providecommand \doibase [0]{http://dx.doi.org/}%
\providecommand \selectlanguage [0]{\@gobble}%
\providecommand \bibinfo  [0]{\@secondoftwo}%
\providecommand \bibfield  [0]{\@secondoftwo}%
\providecommand \translation [1]{[#1]}%
\providecommand \BibitemOpen [0]{}%
\providecommand \bibitemStop [0]{}%
\providecommand \bibitemNoStop [0]{.\EOS\space}%
\providecommand \EOS [0]{\spacefactor3000\relax}%
\providecommand \BibitemShut  [1]{\csname bibitem#1\endcsname}%
\let\auto@bib@innerbib\@empty
\bibitem [{\citenamefont {Szabo}\ and\ \citenamefont
  {Ostlund}(2012)}]{szabo2012modern}%
  \BibitemOpen
  \bibfield  {author} {\bibinfo {author} {\bibfnamefont {Attila}\ \bibnamefont
  {Szabo}}\ and\ \bibinfo {author} {\bibfnamefont {Neil~S}\ \bibnamefont
  {Ostlund}},\ }\href@noop {} {\emph {\bibinfo {title} {{Modern quantum
  chemistry: introduction to advanced electronic structure theory}}}}\
  (\bibinfo  {publisher} {Courier Corporation},\ \bibinfo {year}
  {2012})\BibitemShut {NoStop}%
\bibitem [{\citenamefont {Hohenberg}\ and\ \citenamefont
  {Kohn}(1964)}]{Hohenberg1964}%
  \BibitemOpen
  \bibfield  {author} {\bibinfo {author} {\bibfnamefont {P.}~\bibnamefont
  {Hohenberg}}\ and\ \bibinfo {author} {\bibfnamefont {W.}~\bibnamefont
  {Kohn}},\ }\bibfield  {title} {\enquote {\bibinfo {title} {{Inhomogeneous
  electron gas}},}\ }\href {\doibase 10.1103/PhysRev.136.B864} {\bibfield
  {journal} {\bibinfo  {journal} {Phys. Rev.}\ }\textbf {\bibinfo {volume}
  {136}},  \bibinfo {pages} {B864} (\bibinfo {year} {1964})}\BibitemShut
  {NoStop}%
\bibitem [{\citenamefont {Parr}\ and\ \citenamefont {Yang}(1989)}]{Parr1989}%
  \BibitemOpen
  \bibfield  {author} {\bibinfo {author} {\bibfnamefont {Robert~G}\
  \bibnamefont {Parr}}\ and\ \bibinfo {author} {\bibfnamefont {Weitao}\
  \bibnamefont {Yang}},\ }\href@noop {} {\emph {\bibinfo {title}
  {{Density-Functional Theory of Atoms and Molecules}}}}\ (\bibinfo
  {publisher} {Oxford University Press, New York},\ \bibinfo {year}
  {1989})\BibitemShut {NoStop}%
\bibitem [{\citenamefont {Tsipis}(2014)}]{Tsipis2014}%
  \BibitemOpen
  \bibfield  {author} {\bibinfo {author} {\bibfnamefont {Athanassios~C.}\
  \bibnamefont {Tsipis}},\ }\bibfield  {title} {\enquote {\bibinfo {title}
  {{DFT flavor of coordination chemistry}},}\ }\href {\doibase
  10.1016/j.ccr.2014.02.023} {\bibfield  {journal} {\bibinfo  {journal}
  {Coord. Chem. Rev.}\ }\textbf {\bibinfo {volume} {272}},\
  \bibinfo {pages} {1--29} (\bibinfo {year} {2014})}\BibitemShut {NoStop}%
\bibitem [{\citenamefont {Head-Gordon}(1996)}]{headGordon}%
  \BibitemOpen
  \bibfield  {author} {\bibinfo {author} {\bibfnamefont {Martin}\ \bibnamefont
  {Head-Gordon}},\ }\bibfield  {title} {\enquote {\bibinfo {title} {{Quantum
  Chemistry and Molecular Processes}},}\ }\href {\doibase 10.1021/jp953665+}
  {\bibfield  {journal} {\bibinfo  {journal} {J. Phys. Chem.}\ }\textbf {\bibinfo {volume} {100}},\ \bibinfo {pages}
  {13213--13225} (\bibinfo {year} {1996})}\BibitemShut {NoStop}%
\bibitem [{\citenamefont {Alexandrova}\ \emph {et~al.}(2006)\citenamefont
  {Alexandrova}, \citenamefont {Boldyrev}, \citenamefont {Zhai},\ and\
  \citenamefont {Wang}}]{Alexandrova2006}%
  \BibitemOpen
  \bibfield  {author} {\bibinfo {author} {\bibfnamefont {Anastassia~N.}\
  \bibnamefont {Alexandrova}}, \bibinfo {author} {\bibfnamefont {Alexander~I.}\
  \bibnamefont {Boldyrev}}, \bibinfo {author} {\bibfnamefont {Hua~Jin}\
  \bibnamefont {Zhai}}, \ and\ \bibinfo {author} {\bibfnamefont {Lai~Sheng}\
  \bibnamefont {Wang}},\ }\bibfield  {title} {\enquote {\bibinfo {title}
  {{All-boron aromatic clusters as potential new inorganic ligands and building
  blocks in chemistry}},}\ }\href {\doibase 10.1016/j.ccr.2006.03.032}
  {\bibfield  {journal} {\bibinfo  {journal} {Coord. Chem. Rev.}\
  }\textbf {\bibinfo {volume} {250}},\ \bibinfo {pages} {2811--2866} (\bibinfo
  {year} {2006})}\BibitemShut {NoStop}%
\bibitem [{\citenamefont {Domingo}\ \emph {et~al.}(2016)\citenamefont
  {Domingo}, \citenamefont {R{\'{i}}os-Guti{\'{e}}rrez},\ and\ \citenamefont
  {P{\'{e}}rez}}]{Domingo2016}%
  \BibitemOpen
  \bibfield  {author} {\bibinfo {author} {\bibfnamefont {Luis}\ \bibnamefont
  {Domingo}}, \bibinfo {author} {\bibfnamefont {Mar}\ \bibnamefont
  {R{\'{i}}os-Guti{\'{e}}rrez}}, \ and\ \bibinfo {author} {\bibfnamefont
  {Patricia}\ \bibnamefont {P{\'{e}}rez}},\ }\bibfield  {title} {\enquote
  {\bibinfo {title} {{Applications of the Conceptual Density Functional Theory
  Indices to Organic Chemistry Reactivity}},}\ }\href {\doibase
  10.3390/molecules21060748} {\bibfield  {journal} {\bibinfo  {journal}
  {Molecules}\ }\textbf {\bibinfo {volume} {21}},\ \bibinfo {pages} {748}
  (\bibinfo {year} {2016})}\BibitemShut {NoStop}%
\bibitem [{\citenamefont {Gross}\ and\ \citenamefont {Kohn}(1990)}]{Gross1990}%
  \BibitemOpen
  \bibfield  {author} {\bibinfo {author} {\bibfnamefont {E.~K.U.}\ \bibnamefont
  {Gross}}\ and\ \bibinfo {author} {\bibfnamefont {W.}~\bibnamefont {Kohn}},\
  }\bibfield  {title} {\enquote {\bibinfo {title} {{Time-dependent
  density-functional theory}},}\ }\href {\doibase
  10.1016/S0065-3276(08)60600-0} {\bibfield  {journal} {\bibinfo  {journal}
  {Adv. Quantum Chem.}\ }\textbf {\bibinfo {volume} {21}},\ \bibinfo
  {pages} {255--291} (\bibinfo {year} {1990})}\BibitemShut {NoStop}%
\bibitem [{\citenamefont {White}(1992)}]{White1992}%
  \BibitemOpen
  \bibfield  {author} {\bibinfo {author} {\bibfnamefont {Steven~R.}\
  \bibnamefont {White}},\ }\bibfield  {title} {\enquote {\bibinfo {title}
  {{Density matrix formulation for quantum renormalization groups}},}\ }\href
  {\doibase 10.1103/PhysRevLett.69.2863} {\bibfield  {journal} {\bibinfo
  {journal} {Phys. Rev. Lett.}\ }\textbf {\bibinfo {volume} {69}},\
  \bibinfo {pages} {2863--2866} (\bibinfo {year} {1992})}\BibitemShut {NoStop}%
\bibitem [{\citenamefont {Yang}\ and\ \citenamefont {White}(2019)}]{Yang2019}%
  \BibitemOpen
  \bibfield  {author} {\bibinfo {author} {\bibfnamefont {Mingru}\ \bibnamefont
  {Yang}}\ and\ \bibinfo {author} {\bibfnamefont {Steven~R.}\ \bibnamefont
  {White}},\ }\bibfield  {title} {\enquote {\bibinfo {title}
  {{Density-matrix-renormalization-group study of a one-dimensional diatomic
  molecule beyond the Born-Oppenheimer approximation}},}\ }\href {\doibase
  10.1103/PhysRevA.99.022509} {\bibfield  {journal} {\bibinfo  {journal}
  {Phys. Rev. A}\ }\textbf {\bibinfo {volume} {99}}, \bibinfo {pages} {022509} (\bibinfo {year}
  {2019})}\BibitemShut {NoStop}%
\bibitem [{\citenamefont {Motta}\ \emph {et~al.}(2019)\citenamefont {Motta},
  \citenamefont {Genovese}, \citenamefont {Ma}, \citenamefont {Cui},
  \citenamefont {Sawaya}, \citenamefont {Chan}, \citenamefont {Chepiga},
  \citenamefont {Helms}, \citenamefont {Jimenez-Hoyos}, \citenamefont {Millis},
  \citenamefont {Ray}, \citenamefont {Ronca}, \citenamefont {Shi},
  \citenamefont {Sorella}, \citenamefont {Stoudenmire}, \citenamefont {White},\
  and\ \citenamefont {Zhang}}]{Motta2019}%
  \BibitemOpen
  \bibfield  {author} {\bibinfo {author} {\bibfnamefont {Mario}\ \bibnamefont
  {Motta}}, \bibinfo {author} {\bibfnamefont {Claudio}\ \bibnamefont
  {Genovese}}, \bibinfo {author} {\bibfnamefont {Fengjie}\ \bibnamefont {Ma}},
  \bibinfo {author} {\bibfnamefont {Zhi-Hao}\ \bibnamefont {Cui}}, \bibinfo
  {author} {\bibfnamefont {Randy}\ \bibnamefont {Sawaya}}, \bibinfo {author}
  {\bibfnamefont {Garnet Kin-Lic}\ \bibnamefont {Chan}}, \bibinfo {author}
  {\bibfnamefont {Natalia}\ \bibnamefont {Chepiga}}, \bibinfo {author}
  {\bibfnamefont {Phillip}\ \bibnamefont {Helms}}, \bibinfo {author}
  {\bibfnamefont {Carlos}\ \bibnamefont {Jimenez-Hoyos}}, \bibinfo {author}
  {\bibfnamefont {Andrew~J.}\ \bibnamefont {Millis}}, \bibinfo {author}
  {\bibfnamefont {Ushnish}\ \bibnamefont {Ray}}, \bibinfo {author}
  {\bibfnamefont {Enrico}\ \bibnamefont {Ronca}}, \bibinfo {author}
  {\bibfnamefont {Hao}\ \bibnamefont {Shi}}, \bibinfo {author} {\bibfnamefont
  {Sandro}\ \bibnamefont {Sorella}}, \bibinfo {author} {\bibfnamefont
  {Edwin~M.}\ \bibnamefont {Stoudenmire}}, \bibinfo {author} {\bibfnamefont
  {Steven~R.}\ \bibnamefont {White}}, \ and\ \bibinfo {author} {\bibfnamefont
  {Shiwei}\ \bibnamefont {Zhang}},\ }\bibfield  {title} {\enquote {\bibinfo
  {title} {{Ground-state properties of the hydrogen chain: insulator-to-metal
  transition, dimerization, and magnetic phases}},}\ }\href
  {http://arxiv.org/abs/1911.01618} {\ arXiv:1911.01618  (\bibinfo {year} {2019})}\BibitemShut
  {NoStop}%
\bibitem [{\citenamefont {Motta}\ \emph {et~al.}(2017)\citenamefont {Motta},
  \citenamefont {Ceperley}, \citenamefont {Chan}, \citenamefont {Gomez},
  \citenamefont {Gull}, \citenamefont {Guo}, \citenamefont
  {Jim{\'{e}}nez-Hoyos}, \citenamefont {Lan}, \citenamefont {Li}, \citenamefont
  {Ma}, \citenamefont {Millis}, \citenamefont {Prokof’ev}, \citenamefont
  {Ray}, \citenamefont {Scuseria}, \citenamefont {Sorella}, \citenamefont
  {Stoudenmire}, \citenamefont {Sun}, \citenamefont {Tupitsyn}, \citenamefont
  {White}, \citenamefont {Zgid},\ and\ \citenamefont {Zhang}}]{Motta2017}%
  \BibitemOpen
  \bibfield  {author} {\bibinfo {author} {\bibfnamefont {Mario}\ \bibnamefont
  {Motta}}, \bibinfo {author} {\bibfnamefont {David~M.}\ \bibnamefont
  {Ceperley}}, \bibinfo {author} {\bibfnamefont {Garnet Kin~Lic}\ \bibnamefont
  {Chan}}, \bibinfo {author} {\bibfnamefont {John~A.}\ \bibnamefont {Gomez}},
  \bibinfo {author} {\bibfnamefont {Emanuel}\ \bibnamefont {Gull}}, \bibinfo
  {author} {\bibfnamefont {S.}~\bibnamefont {Guo}}, \bibinfo {author}
  {\bibfnamefont {Carlos~A.}\ \bibnamefont {Jim{\'{e}}nez-Hoyos}}, \bibinfo
  {author} {\bibfnamefont {Tran~Nguyen}\ \bibnamefont {Lan}}, \bibinfo {author}
  {\bibfnamefont {Jia}\ \bibnamefont {Li}}, \bibinfo {author} {\bibfnamefont
  {Fengjie}\ \bibnamefont {Ma}}, \bibinfo {author} {\bibfnamefont {Andrew~J.}\
  \bibnamefont {Millis}}, \bibinfo {author} {\bibfnamefont {Nikolay~V.}\
  \bibnamefont {Prokof’ev}}, \bibinfo {author} {\bibfnamefont {Ushnish}\
  \bibnamefont {Ray}}, \bibinfo {author} {\bibfnamefont {Gustavo~E.}\
  \bibnamefont {Scuseria}}, \bibinfo {author} {\bibfnamefont {Sandro}\
  \bibnamefont {Sorella}}, \bibinfo {author} {\bibfnamefont {Edwin~M.}\
  \bibnamefont {Stoudenmire}}, \bibinfo {author} {\bibfnamefont {Qiming}\
  \bibnamefont {Sun}}, \bibinfo {author} {\bibfnamefont {Igor~S.}\ \bibnamefont
  {Tupitsyn}}, \bibinfo {author} {\bibfnamefont {Steven~R.}\ \bibnamefont
  {White}}, \bibinfo {author} {\bibfnamefont {Dominika}\ \bibnamefont {Zgid}},
  \ and\ \bibinfo {author} {\bibfnamefont {Shiwei}\ \bibnamefont {Zhang}},\
  }\bibfield  {title} {\enquote {\bibinfo {title} {{Towards the solution of the
  many-electron problem in real materials: Equation of state of the hydrogen
  chain with state-of-the-art many-body methods}},}\ }\href {\doibase
  10.1103/PhysRevX.7.031059} {\bibfield  {journal} {\bibinfo  {journal}
  {Phys. Rev. X}\ }\textbf {\bibinfo {volume} {7}} (\bibinfo {year}
  {2017}),\ 10.1103/PhysRevX.7.031059}\BibitemShut {NoStop}%
\bibitem [{\citenamefont {Lubasch}\ \emph {et~al.}(2016)\citenamefont
  {Lubasch}, \citenamefont {Fuks}, \citenamefont {Appel}, \citenamefont
  {Rubio}, \citenamefont {Cirac},\ and\ \citenamefont
  {Ba{\~{n}}uls}}]{Lubasch2016}%
  \BibitemOpen
  \bibfield  {author} {\bibinfo {author} {\bibfnamefont {Michael}\ \bibnamefont
  {Lubasch}}, \bibinfo {author} {\bibfnamefont {Johanna~I.}\ \bibnamefont
  {Fuks}}, \bibinfo {author} {\bibfnamefont {Heiko}\ \bibnamefont {Appel}},
  \bibinfo {author} {\bibfnamefont {Angel}\ \bibnamefont {Rubio}}, \bibinfo
  {author} {\bibfnamefont {J.~Ignacio}\ \bibnamefont {Cirac}}, \ and\ \bibinfo
  {author} {\bibfnamefont {Mari~Carmen}\ \bibnamefont {Ba{\~{n}}uls}},\
  }\bibfield  {title} {\enquote {\bibinfo {title} {{Systematic construction of
  density functionals based on matrix product state computations}},}\ }\href
  {\doibase 10.1088/1367-2630/18/8/083039} {\bibfield  {journal} {\bibinfo
  {journal} {New J. Phys.}\ }\textbf {\bibinfo {volume} {18}}, \bibinfo
  {pages} {083039}
  (\bibinfo {year} {2016})}\BibitemShut
  {NoStop}%
\bibitem [{\citenamefont {Cao}\ \emph {et~al.}(2019)\citenamefont {Cao},
  \citenamefont {Romero}, \citenamefont {Olson}, \citenamefont {Degroote},
  \citenamefont {Johnson}, \citenamefont {Kieferov{\'{a}}}, \citenamefont
  {Kivlichan}, \citenamefont {Menke}, \citenamefont {Peropadre}, \citenamefont
  {Sawaya}, \citenamefont {Sim}, \citenamefont {Veis},\ and\ \citenamefont
  {Aspuru-Guzik}}]{Cao2019}%
  \BibitemOpen
  \bibfield  {author} {\bibinfo {author} {\bibfnamefont {Yudong}\ \bibnamefont
  {Cao}}, \bibinfo {author} {\bibfnamefont {Jonathan}\ \bibnamefont {Romero}},
  \bibinfo {author} {\bibfnamefont {Jonathan~P.}\ \bibnamefont {Olson}},
  \bibinfo {author} {\bibfnamefont {Matthias}\ \bibnamefont {Degroote}},
  \bibinfo {author} {\bibfnamefont {Peter~D.}\ \bibnamefont {Johnson}},
  \bibinfo {author} {\bibfnamefont {Mária}\ \bibnamefont {Kieferov{\'{a}}}},
  \bibinfo {author} {\bibfnamefont {Ian~D.}\ \bibnamefont {Kivlichan}},
  \bibinfo {author} {\bibfnamefont {Tim}\ \bibnamefont {Menke}}, \bibinfo
  {author} {\bibfnamefont {Borja}\ \bibnamefont {Peropadre}}, \bibinfo {author}
  {\bibfnamefont {Nicolas P.~D.}\ \bibnamefont {Sawaya}}, \bibinfo {author}
  {\bibfnamefont {Sukin}\ \bibnamefont {Sim}}, \bibinfo {author} {\bibfnamefont
  {Libor}\ \bibnamefont {Veis}}, \ and\ \bibinfo {author} {\bibfnamefont
  {Alán}\ \bibnamefont {Aspuru-Guzik}},\ }\bibfield  {title} {\enquote
  {\bibinfo {title} {{Quantum Chemistry in the Age of Quantum Computing}},}\
  }\href {\doibase 10.1021/acs.chemrev.8b00803} {\bibfield  {journal} {\bibinfo
   {journal} {Chem. Rev.}\ }\textbf {\bibinfo {volume} {119}},\ \bibinfo
  {pages} {10856--10915} (\bibinfo {year} {2019})}\BibitemShut {NoStop}%
\bibitem [{\citenamefont {Aspuru}\ and\ \citenamefont
  {Guzik}(2005)}]{Aspuru-2005}%
  \BibitemOpen
  \bibfield  {author} {\bibinfo {author} {\bibfnamefont {Citation}\
  \bibnamefont {Aspuru}}\ and\ \bibinfo {author} {\bibfnamefont
  {A}~\bibnamefont {Guzik}},\ }\bibfield  {title} {\enquote {\bibinfo {title}
  {{Simulated Quantum Computation of Molecular Energies}},}\ }\href {\doibase
  10.1126/science.1113479} {\bibfield  {journal} {\bibinfo  {journal}
  {Science}\ }\textbf {\bibinfo {volume} {309}},\ \bibinfo {pages} {1704--1707}
  (\bibinfo {year} {2005})}\BibitemShut {NoStop}%
\bibitem [{\citenamefont {Lanyon}\ \emph {et~al.}(2010)\citenamefont {Lanyon},
  \citenamefont {Whitfield}, \citenamefont {Gillett}, \citenamefont {Goggin},
  \citenamefont {Almeida}, \citenamefont {Kassal}, \citenamefont {Biamonte},
  \citenamefont {Mohseni}, \citenamefont {Powell}, \citenamefont {Barbieri},
  \citenamefont {Aspuru-Guzik},\ and\ \citenamefont {White}}]{Lanyon2010}%
  \BibitemOpen
  \bibfield  {author} {\bibinfo {author} {\bibfnamefont {B.~P.}\ \bibnamefont
  {Lanyon}}, \bibinfo {author} {\bibfnamefont {J.~D.}\ \bibnamefont
  {Whitfield}}, \bibinfo {author} {\bibfnamefont {G.~G.}\ \bibnamefont
  {Gillett}}, \bibinfo {author} {\bibfnamefont {M.~E.}\ \bibnamefont {Goggin}},
  \bibinfo {author} {\bibfnamefont {M.~P.}\ \bibnamefont {Almeida}}, \bibinfo
  {author} {\bibfnamefont {I.}~\bibnamefont {Kassal}}, \bibinfo {author}
  {\bibfnamefont {J.~D.}\ \bibnamefont {Biamonte}}, \bibinfo {author}
  {\bibfnamefont {M.}~\bibnamefont {Mohseni}}, \bibinfo {author} {\bibfnamefont
  {B.~J.}\ \bibnamefont {Powell}}, \bibinfo {author} {\bibfnamefont
  {M.}~\bibnamefont {Barbieri}}, \bibinfo {author} {\bibfnamefont
  {A.}~\bibnamefont {Aspuru-Guzik}}, \ and\ \bibinfo {author} {\bibfnamefont
  {A.~G.}\ \bibnamefont {White}},\ }\bibfield  {title} {\enquote {\bibinfo
  {title} {{Towards quantum chemistry on a quantum computer}},}\ }\href
  {\doibase 10.1038/nchem.483} {\bibfield  {journal} {\bibinfo  {journal}
  {Nat. Chem.}\ }\textbf {\bibinfo {volume} {2}},\ \bibinfo {pages}
  {106--111} (\bibinfo {year} {2010})}\BibitemShut {NoStop}%
\bibitem [{\citenamefont {Kassal}\ \emph {et~al.}(2011)\citenamefont {Kassal},
  \citenamefont {Whitfield}, \citenamefont {Perdomo-Ortiz}, \citenamefont
  {Yung},\ and\ \citenamefont {Aspuru-Guzik}}]{Kassal2011}%
  \BibitemOpen
  \bibfield  {author} {\bibinfo {author} {\bibfnamefont {Ivan}\ \bibnamefont
  {Kassal}}, \bibinfo {author} {\bibfnamefont {James~D}\ \bibnamefont
  {Whitfield}}, \bibinfo {author} {\bibfnamefont {Alejandro}\ \bibnamefont
  {Perdomo-Ortiz}}, \bibinfo {author} {\bibfnamefont {Man-Hong}\ \bibnamefont
  {Yung}}, \ and\ \bibinfo {author} {\bibfnamefont {Alán}\ \bibnamefont
  {Aspuru-Guzik}},\ }\bibfield  {title} {\enquote {\bibinfo {title}
  {{Simulating chemistry using quantum computers}},}\ }\href {\doibase
  10.1146/annurev-physchem-032210-103512} {\bibfield  {journal} {\bibinfo
  {journal} {Annu. Rev. Phys. Chem.}\ }\textbf {\bibinfo {volume} {62}},\
  \bibinfo {pages} {185--207} (\bibinfo {year} {2011})}\BibitemShut {NoStop}%
\bibitem [{\citenamefont {Wecker}\ \emph {et~al.}(2015)\citenamefont {Wecker},
  \citenamefont {Hastings},\ and\ \citenamefont {Troyer}}]{Wecker2015}%
  \BibitemOpen
  \bibfield  {author} {\bibinfo {author} {\bibfnamefont {Dave}\ \bibnamefont
  {Wecker}}, \bibinfo {author} {\bibfnamefont {Matthew~B.}\ \bibnamefont
  {Hastings}}, \ and\ \bibinfo {author} {\bibfnamefont {Matthias}\ \bibnamefont
  {Troyer}},\ }\bibfield  {title} {\enquote {\bibinfo {title} {{Progress
  towards practical quantum variational algorithms}},}\ }\href {\doibase
  10.1103/PhysRevA.92.042303} {\bibfield  {journal} {\bibinfo  {journal}
  {Phys. Rev. A}\ }\textbf {\bibinfo {volume} {92}}, \bibinfo {pages} {042303} (\bibinfo {year}
  {2015})}\BibitemShut {NoStop}%
\bibitem [{\citenamefont {Higgott}\ \emph {et~al.}(2019)\citenamefont
  {Higgott}, \citenamefont {Wang},\ and\ \citenamefont
  {Brierley}}]{Higgott2019}%
  \BibitemOpen
  \bibfield  {author} {\bibinfo {author} {\bibfnamefont {Oscar}\ \bibnamefont
  {Higgott}}, \bibinfo {author} {\bibfnamefont {Daochen}\ \bibnamefont {Wang}},
  \ and\ \bibinfo {author} {\bibfnamefont {Stephen}\ \bibnamefont {Brierley}},\
  }\bibfield  {title} {\enquote {\bibinfo {title} {{Variational Quantum
  Computation of Excited States}},}\ }\href {\doibase
  10.22331/q-2019-07-01-156} {\bibfield  {journal} {\bibinfo  {journal}
  {Quantum}\ }\textbf {\bibinfo {volume} {3}},\ \bibinfo {pages} {156}
  (\bibinfo {year} {2019})}\BibitemShut {NoStop}%
\bibitem [{\citenamefont {Arg{\"{u}}ello-Luengo}\ \emph
  {et~al.}(2019)\citenamefont {Arg{\"{u}}ello-Luengo}, \citenamefont
  {Gonz{\'{a}}lez-Tudela}, \citenamefont {Shi}, \citenamefont {Zoller},\ and\
  \citenamefont {Cirac}}]{arguello2019analogue}%
  \BibitemOpen
  \bibfield  {author} {\bibinfo {author} {\bibfnamefont {Javier}\ \bibnamefont
  {Arg{\"{u}}ello-Luengo}}, \bibinfo {author} {\bibfnamefont {Alejandro}\
  \bibnamefont {Gonz{\'{a}}lez-Tudela}}, \bibinfo {author} {\bibfnamefont
  {Tao}\ \bibnamefont {Shi}}, \bibinfo {author} {\bibfnamefont {Peter}\
  \bibnamefont {Zoller}}, \ and\ \bibinfo {author} {\bibfnamefont {J.~Ignacio}\
  \bibnamefont {Cirac}},\ }\bibfield  {title} {\enquote {\bibinfo {title}
  {{Analogue quantum chemistry simulation}},}\ }\href {\doibase
  10.1038/s41586-019-1614-4} {\bibfield  {journal} {\bibinfo  {journal}
  {Nature}\ }\textbf {\bibinfo {volume} {574}},\ \bibinfo {pages} {215--218}
  (\bibinfo {year} {2019})}\BibitemShut {NoStop}%
\bibitem [{\citenamefont {Bloch}\ \emph {et~al.}(2008)\citenamefont {Bloch},
  \citenamefont {Dalibard},\ and\ \citenamefont {Zwerger}}]{bloch08a}%
  \BibitemOpen
  \bibfield  {author} {\bibinfo {author} {\bibfnamefont {Immanuel}\
  \bibnamefont {Bloch}}, \bibinfo {author} {\bibfnamefont {Jean}\ \bibnamefont
  {Dalibard}}, \ and\ \bibinfo {author} {\bibfnamefont {Wilhelm}\ \bibnamefont
  {Zwerger}},\ }\bibfield  {title} {\enquote {\bibinfo {title} {{Many-body
  physics with ultracold gases}},}\ }\href {\doibase 10.1103/RevModPhys.80.885}
  {\bibfield  {journal} {\bibinfo  {journal} {Rev. Mod. Phys.}\
  }\textbf {\bibinfo {volume} {80}},\ \bibinfo {pages} {885--964} (\bibinfo
  {year} {2008})}\BibitemShut {NoStop}%
\bibitem [{\citenamefont {Esslinger}(2010)}]{Esslinger2010}%
  \BibitemOpen
  \bibfield  {author} {\bibinfo {author} {\bibfnamefont {Tilman}\ \bibnamefont
  {Esslinger}},\ }\bibfield  {title} {\enquote {\bibinfo {title}
  {{Fermi-Hubbard Physics with Atoms in an Optical Lattice}},}\ }\href
  {\doibase 10.1146/annurev-conmatphys-070909-104059} {\bibfield  {journal}
  {\bibinfo  {journal} {Annu. Rev. Condens. Matter Phys.}\ }\textbf
  {\bibinfo {volume} {1}},\ \bibinfo {pages} {129--152} (\bibinfo {year}
  {2010})}\BibitemShut {NoStop}%
\bibitem [{\citenamefont {Gross}\ and\ \citenamefont
  {Bloch}(2017)}]{Gross2017}%
  \BibitemOpen
  \bibfield  {author} {\bibinfo {author} {\bibfnamefont {Christian}\
  \bibnamefont {Gross}}\ and\ \bibinfo {author} {\bibfnamefont {Immanuel}\
  \bibnamefont {Bloch}},\ }\bibfield  {title} {\enquote {\bibinfo {title}
  {{Quantum simulations with ultracold atoms in optical lattices.}}}\ }\href
  {\doibase 10.1126/science.aal3837} {\bibfield  {journal} {\bibinfo  {journal}
  {Science}\ }\textbf {\bibinfo {volume} {357}},\ \bibinfo
  {pages} {995--1001} (\bibinfo {year} {2017})}\BibitemShut {NoStop}%
\bibitem [{Note1()}]{Note1}%
  \BibitemOpen
  \bibinfo {note} {We acknowledge that other analog simulators based on
  fermionic atoms trapped in optical lattices have been proposed to emulate the
  molecular potentials of benzene-like molecules~\cite {Luhmann2015} or
  simulate ultrafast dynamics in strong-fields~\cite {Sala2017,Senaratne2018}.
  In contrast to them, Ref.~\cite {arguello2019analogue} and the present
  proposal allow to go beyond the local interactions naturally found in cold
  atoms, simulating the non-local fermionic repulsion that appears in typical
  quantum chemistry problems.}\BibitemShut {Stop}%
  \bibitem [{\citenamefont {L{\"{u}}hmann}\ \emph {et~al.}(2015)\citenamefont
  {L{\"{u}}hmann}, \citenamefont {Weitenberg},\ and\ \citenamefont
  {Sengstock}}]{Luhmann2015}%
  \BibitemOpen
  \bibfield  {author} {\bibinfo {author} {\bibfnamefont {Dirk-Sören}\
  \bibnamefont {L{\"{u}}hmann}}, \bibinfo {author} {\bibfnamefont {Christof}\
  \bibnamefont {Weitenberg}}, \ and\ \bibinfo {author} {\bibfnamefont {Klaus}\
  \bibnamefont {Sengstock}},\ }\bibfield  {title} {\enquote {\bibinfo {title}
  {{Emulating Molecular Orbitals and Electronic Dynamics with Ultracold
  Atoms}},}\ }\href {\doibase 10.1103/PhysRevX.5.031016} {\bibfield  {journal}
  {\bibinfo  {journal} {Phys. Rev. X}\ }\textbf {\bibinfo {volume} {5}},\
  \bibinfo {pages} {031016} (\bibinfo {year} {2015})}\BibitemShut {NoStop}%
\bibitem [{\citenamefont {Sala}\ \emph {et~al.}(2017)\citenamefont {Sala},
  \citenamefont {F{\"{o}}rster},\ and\ \citenamefont {Saenz}}]{Sala2017}%
  \BibitemOpen
  \bibfield  {author} {\bibinfo {author} {\bibfnamefont {Simon}\ \bibnamefont
  {Sala}}, \bibinfo {author} {\bibfnamefont {Johann}\ \bibnamefont
  {F{\"{o}}rster}}, \ and\ \bibinfo {author} {\bibfnamefont {Alejandro}\
  \bibnamefont {Saenz}},\ }\bibfield  {title} {\enquote {\bibinfo {title}
  {{Ultracold-atom quantum simulator for attosecond science}},}\ }\href
  {\doibase 10.1103/PhysRevA.95.011403} {\bibfield  {journal} {\bibinfo
  {journal} {Phys. Rev. A}\ }\textbf {\bibinfo {volume} {95}},\ \bibinfo
  {pages} {11403} (\bibinfo {year} {2017})}\BibitemShut {NoStop}%
\bibitem [{\citenamefont {Senaratne}\ \emph {et~al.}(2018)\citenamefont
  {Senaratne}, \citenamefont {Rajagopal}, \citenamefont {Shimasaki},
  \citenamefont {Dotti}, \citenamefont {Fujiwara}, \citenamefont {Singh},
  \citenamefont {Geiger},\ and\ \citenamefont {Weld}}]{Senaratne2018}%
  \BibitemOpen
  \bibfield  {author} {\bibinfo {author} {\bibfnamefont {Ruwan}\ \bibnamefont
  {Senaratne}}, \bibinfo {author} {\bibfnamefont {Shankari~V.}\ \bibnamefont
  {Rajagopal}}, \bibinfo {author} {\bibfnamefont {Toshihiko}\ \bibnamefont
  {Shimasaki}}, \bibinfo {author} {\bibfnamefont {Peter~E.}\ \bibnamefont
  {Dotti}}, \bibinfo {author} {\bibfnamefont {Kurt~M.}\ \bibnamefont
  {Fujiwara}}, \bibinfo {author} {\bibfnamefont {Kevin}\ \bibnamefont {Singh}},
  \bibinfo {author} {\bibfnamefont {Zachary~A.}\ \bibnamefont {Geiger}}, \ and\
  \bibinfo {author} {\bibfnamefont {David~M.}\ \bibnamefont {Weld}},\
  }\bibfield  {title} {\enquote {\bibinfo {title} {{Quantum simulation of
  ultrafast dynamics using trapped ultracold atoms}},}\ }\href {\doibase
  10.1038/s41467-018-04556-3} {\bibfield  {journal} {\bibinfo  {journal}
  {Nat. Commun.}\ }\textbf {\bibinfo {volume} {9}},\ \bibinfo {pages}
  {2065} (\bibinfo {year} {2018})}\BibitemShut {NoStop}%
\bibitem [{Note2()}]{Note2}%
  \BibitemOpen
  \bibinfo {note} {Throughout the text, bold variables denote 2D
  vectors.}\BibitemShut {Stop}%
\bibitem [{Note3()}]{Note3}%
  \BibitemOpen
  \bibinfo {note} {In order to prevent the divergence in the origin, positions
  $r_n$ of the nuclei are shifted half a site from the lattice nodes in the $y$
  direction.}\BibitemShut {Stop}%
\bibitem [{Note4()}]{Note4}%
  \BibitemOpen
  \bibinfo {note} {Considering that the electronic dynamics is much faster than
  the nuclear one, their equations can be decoupled (Born-Oppenheimer
  approximation). The position $\left \protect \{ \protect \cc@accent
  {"707E}\protect \mathbf {r}_n\right \protect \} _{i=n\protect \ldots N_n}$ of
  the $N_n$ nuclei is considered fixed during the calculation of the electronic
  Hamiltonian $H_\protect \text {cont}$, for the $N_f$ electrons in positions
  $\left \protect \{ \protect \mathbf {r}_i\right \protect \} _{i=1\protect
  \ldots N_f}$. \begin {equation*} \begin {split} H_\protect \text
  {cont}=&-\DOTSB \sum@ \slimits@ _{i=1}^{N_f}\protect \frac {\hbar
  ^2}{2m_e}\nabla ^2_i -\DOTSB \sum@ \slimits@ _{i=1}^{N_f} \protect \frac
  {1}{2}\DOTSB \sum@ \slimits@ _{n=1}^{N_n} Z_n V(\left | \protect \mathbf
  {r}_i-\protect \cc@accent {"707E}\protect \mathbf {r}_n\right |)\\ & + \DOTSB
  \sum@ \slimits@ _{i\not =j=1}^{N_f} V(\left | \protect \mathbf {r}_i-\protect
  \mathbf {r}_j\right |) \protect \tmspace +\thinmuskip {.1667em}, \end {split}
  \end {equation*} where $m_e$ is the mass of the electron and $Z_n$ is the
  atomic number of nucleus $n$. The first term then describes the kinetic
  energy of the electrons, the second its nuclear attraction following the
  potential $V(r)\protect \tmspace +\thinmuskip {.1667em},$ and the third the
  electronic repulsion.}\BibitemShut {Stop}%
\bibitem [{Note5()}]{Note5}%
  \BibitemOpen
  \bibinfo {note} {This externally induced potential could eventually mimic the
  effect of inner-shell electrons as well.}\BibitemShut {Stop}%
\bibitem [{\citenamefont {Choi}\ \emph {et~al.}(2016)\citenamefont {Choi},
  \citenamefont {Hild}, \citenamefont {Zeiher}, \citenamefont {Schau{\ss}},
  \citenamefont {Rubio-Abadal}, \citenamefont {Yefsah}, \citenamefont
  {Khemani}, \citenamefont {Huse}, \citenamefont {Bloch},\ and\ \citenamefont
  {Gross}}]{choi16a}%
  \BibitemOpen
  \bibfield  {author} {\bibinfo {author} {\bibfnamefont {Jae-yoon}\
  \bibnamefont {Choi}}, \bibinfo {author} {\bibfnamefont {Sebastian}\
  \bibnamefont {Hild}}, \bibinfo {author} {\bibfnamefont {Johannes}\
  \bibnamefont {Zeiher}}, \bibinfo {author} {\bibfnamefont {Peter}\
  \bibnamefont {Schau{\ss}}}, \bibinfo {author} {\bibfnamefont {Antonio}\
  \bibnamefont {Rubio-Abadal}}, \bibinfo {author} {\bibfnamefont {Tarik}\
  \bibnamefont {Yefsah}}, \bibinfo {author} {\bibfnamefont {Vedika}\
  \bibnamefont {Khemani}}, \bibinfo {author} {\bibfnamefont {David~A}\
  \bibnamefont {Huse}}, \bibinfo {author} {\bibfnamefont {Immanuel}\
  \bibnamefont {Bloch}}, \ and\ \bibinfo {author} {\bibfnamefont {Christian}\
  \bibnamefont {Gross}},\ }\bibfield  {title} {\enquote {\bibinfo {title}
  {{Exploring the many-body localization transition in two dimensions.}}}\
  }\href {\doibase 10.1126/science.aaf8834} {\bibfield  {journal} {\bibinfo
  {journal} {Science}\ }\textbf {\bibinfo {volume} {352}},\ \bibinfo {pages}
  {1547--52} (\bibinfo {year} {2016})}\BibitemShut {NoStop}%
\bibitem [{\citenamefont {Zaslow}\ and\ \citenamefont
  {Zandler}(1967)}]{Zaslow1967}%
  \BibitemOpen
  \bibfield  {author} {\bibinfo {author} {\bibfnamefont {B}~\bibnamefont
  {Zaslow}}\ and\ \bibinfo {author} {\bibfnamefont {Melvin~E}\ \bibnamefont
  {Zandler}},\ }\bibfield  {title} {\enquote {\bibinfo {title}
  {{Two-Dimensional Analog to the Hydrogen Atom Exact analytical solutions of a
  two-dimensional hydrogen atom in a constant magnetic field}},}\ }\href
  {\doibase 10.1063/1.1503868} {\bibfield  {journal} {\bibinfo  {journal}
  {Am. J. Phys.}\ }\textbf {\bibinfo {volume} {35}},\ \bibinfo
  {pages} {1118--1005} (\bibinfo {year} {1967})}\BibitemShut {NoStop}%
\bibitem [{\citenamefont {Zhu}\ and\ \citenamefont {Xiong}(1990)}]{Zhu1990}%
  \BibitemOpen
  \bibfield  {author} {\bibinfo {author} {\bibfnamefont {Jia-Lin}\ \bibnamefont
  {Zhu}}\ and\ \bibinfo {author} {\bibfnamefont {Jia-Jiong}\ \bibnamefont
  {Xiong}},\ }\bibfield  {title} {\enquote {\bibinfo {title} {{Hydrogen
  molecular ions in two dimensions}},}\ }\href {\doibase
  10.1103/PhysRevB.41.12274} {\bibfield  {journal} {\bibinfo  {journal}
  {Phys. Rev. B}\ }\textbf {\bibinfo {volume} {41}},\ \bibinfo {pages}
  {12274--12277} (\bibinfo {year} {1990})}\BibitemShut {NoStop}%
\bibitem [{Note6()}]{Note6}%
  \BibitemOpen
  \bibinfo {note} {As compared to the three-dimensional case, $\protect \text
  {Ry\protect \tmspace +\thinmuskip {.1667em}(2D)}=\protect \text {4Ry\protect
  \tmspace +\thinmuskip {.1667em}(3D)}$, and $2a_0\protect \tmspace
  +\thinmuskip {.1667em}\protect \text {(2D)}=a_0\protect \tmspace +\thinmuskip
  {.1667em} \protect \text {(3D)}$. Throughout the text, we will omit the (2D)
  labelling.}\BibitemShut {Stop}%
\bibitem [{Note7()}]{Note7}%
  \BibitemOpen
  \bibinfo {note} {See the Supplementary material accompanying this Letter.
  Section A discusses the scaling of the spectrum of the discretized 2D
  Hamiltonian as the lattice size increases. Section B derives the effective
  interaction mediated by a single boson with one long-lived state. Section C
  focuses on the effective interaction mediated by several mediating atoms with
  two long-lived internal states. Section D includes further details about the
  numerical calculations shown in Fig. 2-4.}\BibitemShut {Stop}%
\bibitem [{\citenamefont {Patil}(2003)}]{Patil2003}%
  \BibitemOpen
  \bibfield  {author} {\bibinfo {author} {\bibfnamefont {S.~H.}\ \bibnamefont
  {Patil}},\ }\bibfield  {title} {\enquote {\bibinfo {title} {{Hydrogen
  molecular ion and molecule in two dimensions}},}\ }\href {\doibase
  10.1063/1.1531103} {\bibfield  {journal} {\bibinfo  {journal} {J. Chem. Phys.}\ }\textbf {\bibinfo {volume} {118}},\ \bibinfo {pages}
  {2197--2205} (\bibinfo {year} {2003})}\BibitemShut {NoStop}%
\bibitem [{Note8()}]{Note8}%
  \BibitemOpen
  \bibinfo {note} {Note that this choice of nuclear potential differs from the
  one encountered in a flatland world, in which Coulomb's law leads to
  interactions that scale as $\propto \protect \qopname \relax
  o{log}(r)$.}\BibitemShut {Stop}%
\bibitem [{\citenamefont {Lukin}\ \emph {et~al.}(2000)\citenamefont {Lukin},
  \citenamefont {Fleischhauer}, \citenamefont {Cote}, \citenamefont {Duan},
  \citenamefont {Jaksch}, \citenamefont {Cirac},\ and\ \citenamefont
  {Zoller}}]{Lukin2000}%
  \BibitemOpen
  \bibfield  {author} {\bibinfo {author} {\bibfnamefont {M.~D.}\ \bibnamefont
  {Lukin}}, \bibinfo {author} {\bibfnamefont {M.}~\bibnamefont {Fleischhauer}},
  \bibinfo {author} {\bibfnamefont {R.}~\bibnamefont {Cote}}, \bibinfo {author}
  {\bibfnamefont {L.~M.}\ \bibnamefont {Duan}}, \bibinfo {author}
  {\bibfnamefont {D.}~\bibnamefont {Jaksch}}, \bibinfo {author} {\bibfnamefont
  {J.~I.}\ \bibnamefont {Cirac}}, \ and\ \bibinfo {author} {\bibfnamefont
  {P.}~\bibnamefont {Zoller}},\ }\bibfield  {title} {\enquote {\bibinfo {title}
  {{Dipole Blockade and Quantum Information Processing in Mesoscopic Atomic
  Ensembles}},}\ }\href {\doibase 10.1103/PhysRevLett.87.037901},  \bibinfo {pages} {037901} {\  (\bibinfo
  {year} {2000})}\BibitemShut {NoStop}%
\bibitem [{\citenamefont {Ravets}\ \emph {et~al.}(2014)\citenamefont {Ravets},
  \citenamefont {Labuhn}, \citenamefont {Barredo}, \citenamefont
  {B{\'{e}}guin}, \citenamefont {Lahaye},\ and\ \citenamefont
  {Browaeys}}]{Ravets2014}%
  \BibitemOpen
  \bibfield  {author} {\bibinfo {author} {\bibfnamefont {Sylvain}\ \bibnamefont
  {Ravets}}, \bibinfo {author} {\bibfnamefont {Henning}\ \bibnamefont
  {Labuhn}}, \bibinfo {author} {\bibfnamefont {Daniel}\ \bibnamefont
  {Barredo}}, \bibinfo {author} {\bibfnamefont {Lucas}\ \bibnamefont
  {B{\'{e}}guin}}, \bibinfo {author} {\bibfnamefont {Thierry}\ \bibnamefont
  {Lahaye}}, \ and\ \bibinfo {author} {\bibfnamefont {Antoine}\ \bibnamefont
  {Browaeys}},\ }\bibfield  {title} {\enquote {\bibinfo {title} {{Coherent
  dipole-dipole coupling between two single Rydberg atoms at an
  electrically-tuned F{\"{o}}rster resonance}},}\ }\href {\doibase
  10.1038/nphys3119} {\bibfield  {journal} {\bibinfo  {journal} {Nat. Phys.}\ }\textbf {\bibinfo {volume} {10}},\ \bibinfo {pages} {914--917}
  (\bibinfo {year} {2014})}\BibitemShut {NoStop}%
\bibitem [{\citenamefont {Saffman}\ \emph {et~al.}(2010)\citenamefont
  {Saffman}, \citenamefont {Walker},\ and\ \citenamefont
  {M{\o}lmer}}]{Saffman2010}%
  \BibitemOpen
  \bibfield  {author} {\bibinfo {author} {\bibfnamefont {M.}~\bibnamefont
  {Saffman}}, \bibinfo {author} {\bibfnamefont {T.~G.}\ \bibnamefont {Walker}},
  \ and\ \bibinfo {author} {\bibfnamefont {K.}~\bibnamefont {M{\o}lmer}},\
  }\bibfield  {title} {\enquote {\bibinfo {title} {{Quantum information with
  Rydberg atoms}},}\ }\href {\doibase 10.1103/RevModPhys.82.2313} {\bibfield
  {journal} {\bibinfo  {journal} {Rev. Mod. Phys.}\ }\textbf
  {\bibinfo {volume} {82}},\ \bibinfo {pages} {2313--2363} (\bibinfo {year}
  {2010})}\BibitemShut {NoStop}%
\bibitem [{Note9()}]{Note9}%
  \BibitemOpen
  \bibinfo {note} {However, the bare interaction is anisotropic in
  nature.}\BibitemShut {Stop}%
\bibitem [{\citenamefont {Heinz}\ \emph {et~al.}(2019)\citenamefont {Heinz},
  \citenamefont {Park}, \citenamefont {{\v{S}}anti{\'{c}}}, \citenamefont
  {Trautmann}, \citenamefont {Porsev}, \citenamefont {Safronova}, \citenamefont
  {Bloch},\ and\ \citenamefont {Blatt}}]{Heinz2019}%
  \BibitemOpen
  \bibfield  {author} {\bibinfo {author} {\bibfnamefont {A.}~\bibnamefont
  {Heinz}}, \bibinfo {author} {\bibfnamefont {A.~J.}\ \bibnamefont {Park}},
  \bibinfo {author} {\bibfnamefont {N.}~\bibnamefont {{\v{S}}anti{\'{c}}}},
  \bibinfo {author} {\bibfnamefont {J.}~\bibnamefont {Trautmann}}, \bibinfo
  {author} {\bibfnamefont {S.~G.}\ \bibnamefont {Porsev}}, \bibinfo {author}
  {\bibfnamefont {M.~S.}\ \bibnamefont {Safronova}}, \bibinfo {author}
  {\bibfnamefont {I.}~\bibnamefont {Bloch}}, \ and\ \bibinfo {author}
  {\bibfnamefont {S.}~\bibnamefont {Blatt}},\ }\bibfield  {title} {\enquote
  {\bibinfo {title} {{State-dependent optical lattices for the strontium
  optical qubit}},}\ }\href {http://arxiv.org/abs/1912.10350} {\ arXiv:1912.10350 (\bibinfo
  {year} {2019})}\BibitemShut {NoStop}%
\bibitem [{\citenamefont {Katsura}\ and\ \citenamefont
  {Inawashiro}(1971)}]{Katsura1971}%
  \BibitemOpen
  \bibfield  {author} {\bibinfo {author} {\bibfnamefont {Shigetoshi}\
  \bibnamefont {Katsura}}\ and\ \bibinfo {author} {\bibfnamefont {Sakari}\
  \bibnamefont {Inawashiro}},\ }\bibfield  {title} {\enquote {\bibinfo {title}
  {{Lattice Green's Functions for the Rectangular and the Square Lattices at
  Arbitrary Points}},}\ }\href {\doibase 10.1063/1.1665785} {\bibfield
  {journal} {\bibinfo  {journal} {J. Math. Phys.}\ }\textbf
  {\bibinfo {volume} {12}},\ \bibinfo {pages} {1622--1630} (\bibinfo {year}
  {1971})}\BibitemShut {NoStop}%
\bibitem [{\citenamefont {Abramowitz}\ and\ \citenamefont
  {Stegun}(1972)}]{abramowitz1972integration}%
  \BibitemOpen
  \bibfield  {author} {\bibinfo {author} {\bibfnamefont {M}~\bibnamefont
  {Abramowitz}}\ and\ \bibinfo {author} {\bibfnamefont {I~A}\ \bibnamefont
  {Stegun}},\ }\bibfield  {title} {\enquote {\bibinfo {title} {{Handbook of
  Mathematical Functions with Formulas, Graphs and Mathematical Tables}},}\
  }\href@noop {} {\bibfield  {journal} {\bibinfo  {journal} {9th printing, New
  York: Dover}\ } (\bibinfo {year} {1972})}\BibitemShut {NoStop}%
\bibitem [{\citenamefont {Schmid}\ \emph {et~al.}(2006)\citenamefont {Schmid},
  \citenamefont {Thalhammer}, \citenamefont {Winkler}, \citenamefont {Lang},\
  and\ \citenamefont {Denschlag}}]{Schmid2006}%
  \BibitemOpen
  \bibfield  {author} {\bibinfo {author} {\bibfnamefont {Stefan}\ \bibnamefont
  {Schmid}}, \bibinfo {author} {\bibfnamefont {Gregor}\ \bibnamefont
  {Thalhammer}}, \bibinfo {author} {\bibfnamefont {Klaus}\ \bibnamefont
  {Winkler}}, \bibinfo {author} {\bibfnamefont {Florian}\ \bibnamefont {Lang}},
  \ and\ \bibinfo {author} {\bibfnamefont {Johannes~Hecker}\ \bibnamefont
  {Denschlag}},\ }\bibfield  {title} {\enquote {\bibinfo {title} {{Long
  distance transport of ultracold atoms using a 1D optical lattice}},}\ }\href
  {\doibase 10.1088/1367-2630/8/8/159} {\bibfield  {journal} {\bibinfo
  {journal} {New J. Phys.}\ }\textbf {\bibinfo {volume} {8}}, \bibinfo {pages} 159
  (\bibinfo {year} {2006})}\BibitemShut {NoStop}%
\end{thebibliography}
\end{document}